\documentclass[num-refs]{nbdt-article}

\pdfoutput=1

\papertype{Original Article}
\paperfield{Analysis}

\title{Nonlinear demixed component analysis for neural population data as a low-rank kernel regression problem}

\author[1\authfn{1}]{Kenneth W. Latimer}

\affil[1]{Department of Neurobiology, The University of Chicago, Chicago, IL, 60637, USA}

\corraddress{Kenneth W. Latimer, Department of
Neurobiology, University of Chicago, 5812 S. Ellis Avenue, MC0912, P-416, Chicago, IL 60637}
\corremail{latimerk@uchicago.edu}

\fundinginfo{This work was supported by a Chicago Fellowship.}

\runningauthor{Latimer}

\newcommand{\figref}[2][]{Fig.~\ref{#2}#1} 
\renewcommand{\eqref}[2][]{Eq.~\ref{#2}#1} 
\newcommand{\eqsref}[2][]{Eqs.~\ref{#2}#1} 
\newcommand{\tableref}[2][]{Table~\ref{#2}#1} 
\newcommand{\mx}{\mathbf{X}}
\newcommand{\mw}{\mathbf{W}}
\newcommand{\ml}{\mathbf{L}}
\newcommand{\mf}{\mathbf{F}}
\newcommand{\md}{\mathbf{D}}
\newcommand{\mc}{\mathbf{C}}
\newcommand{\eye}{\mathbf{I}}
\newcommand{\mg}{\mathbf{G}}
\newcommand{\mh}{\mathbf{H}}

\newcommand{\mV}{\mathbf{V}}
\newcommand{\mz}{\mathbf{Z}}

\newcommand{\mk}{\mathbf{K}}
\newcommand{\vx}{\mathbf{x}}
\newcommand{\vy}{\mathbf{y}}

\begin{document}

\maketitle

\begin{abstract}
Many studies of neural activity in behaving animals aim to discover interpretable low-dimensional structure in large-scale neural population recordings.
One approach to this problem is demixed principal component analysis (dPCA), a supervised linear dimensionality reduction technique to find components that depend on particular experimental parameters.
Here, I introduce kernel dPCA (kdPCA) as a nonlinear extension of dPCA by applying kernel least-squares regression to the demixing problem.
I consider simulated examples of neural populations with low-dimensional activity to compare the components recovered from dPCA and kdPCA.
These simulations demonstrate that neurally relevant nonlinearities, such as stimulus-dependent gain and rotation, interfere with linear demixing of neural activity into components that represent to individual experimental parameters.
However, kdPCA can still recover interpretable components in these examples.
Finally, I demonstrate kdPCA using two examples of neural populations recorded during perceptual decision-making tasks.

\keywords{dimensionality reduction, kernel regression, neural population analysis, gain modulation}
\end{abstract}

\section{Introduction}
Dimensionality reduction techniques have become an essential step in many analyses of large-scale neural recordings~\citep{CunninghamYu2014,GaoGanguli2015,PangLansdellFairhall2016,WilliamsonEtAl2019,TrautmannEtAl2019}.
Recent work on dimensionality reduction has focused on methods to discover components of neural activity which are aligned to experimental variables of interest, such as stimulus condition or behavioral response~\citep{ChurchlandEtAl2012,KobakEtAl2016,WilliamsEtAl2018}.
Here, I focus on one particular method: demixed PCA, introduced by~\cite{KobakEtAl2016}.
A key assumption of this method is that neural activity lies in a low-dimensional space that can be decomposed as a sum of elements that depend only on particular experimental variables.
This is demonstrated by the example in \figref[A]{fig:demo}, where the low-dimensional activity is the sum of stimulus and time components plus a small interaction term.
The time component can be extracted independently of stimulus by taking a linear projection of the 2-D space, and thus time and stimulus can be demixed. 

However, neural responses may reflect nonlinear mixtures of experimental parameters.
For example, the trajectories in the low-dimensional latent space could show condition-dependent scaling (as with stimulus-dependent gain) or rotation while still maintaining separability.
Although dPCA considers interaction components to account for dependencies between variables, dPCA can still fail to recover components that depend solely on particular task parameters (i.e., does not demix activity) in such conditions.
This challenge is illustrated in~\figref[B]{fig:demo}.
In this example, taking the average over stimulus (or time) reveals the true 1-D time (or stimulus) component.
However, the interaction component extends significantly into the same 2-D space spanned by the stimulus and time components.
The interaction reorients and scales the trajectories of each condition within the 2-D space.
As a result, any linear projection of this 2-D space will contain both stimulus and time information: a failure to demix.
However, it may be possible to find a low-dimensional set of {\emph{nonlinear components}} that successfully demix the neural activity contributed by each task parameter.

\begin{figure}[t!]
\centering
\includegraphics[width=0.8\linewidth]{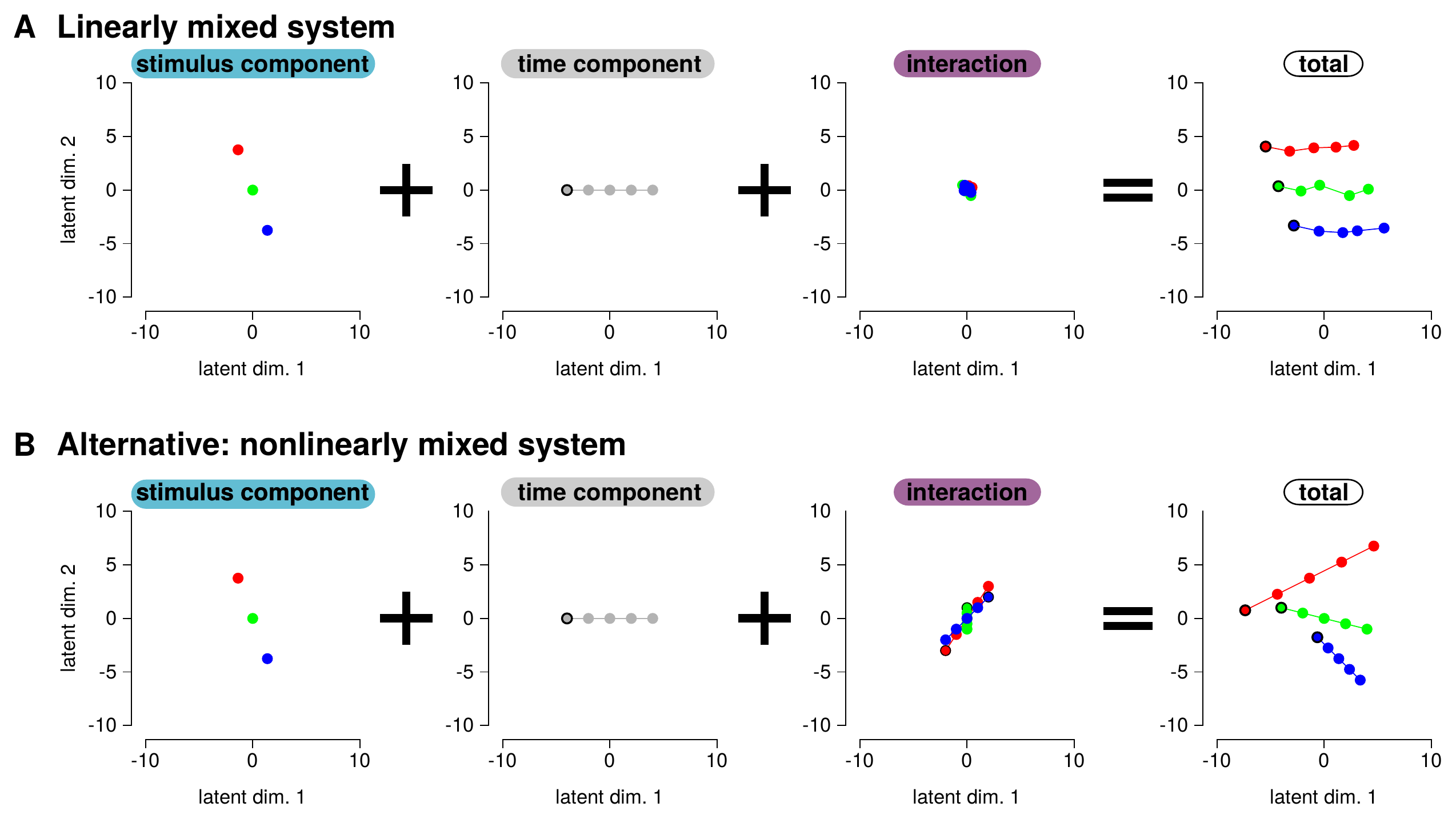}
\caption{Two dimensional subspaces representing latent neural activity.
({\bf A}) In this example, neural activity is the sum of the stimulus component and a time component.
The time and stimulus are both one-dimensional, and together span the 2-D space.
The interaction is between time and stimulus is negligible in the 2-D space.
The right panel shows the total signal. Each colored trace shows the activity as a function of time for a given stimulus.
See similar example in Figure 2 of Kobak {\emph{et al.}}.
Because stimulus and time are still one dimensional and non-overlapping, linear projections can recover the time and stimulus information independent from one another.
({\bf B}) In this example, the time and stimulus components are the same as in A.
An interaction term is introduced that stretches and rotates each condition.
The interaction is constructed so that averaging across time (or stimulus) gives the true stimulus (or time) component.
}
\label{fig:demo}
\end{figure}

I present an extension to dPCA to find nonlinear components that depend on particular task variables.
This method is related to kernel-based extensions of standard principal component analysis ~\citep[kPCA;][]{ScholkopfSmolaMuller1998}, kernel canonical correlation analysis~\citep{LaiFyfe2000,BachJordan2002,HardoonSzedmakShawe2004,RoduEtAl2018}, and kernel regularized least-squares regression~\citep{FriedmanEtAl2001}.
Using a standard kernel approach with low-rank matrix approximations, kdPCA recovers a low-dimensional set of nonlinear components that represents the high-dimensional neural activity while demixing the task variables.
I demonstrate the utility of this method using four simulations and compare to dPCA.
Finally, I apply kdPCA to recover decision-related activity from two datasets recorded during perceptual decision-making tasks.
The first dataset consists of a population of neurons recorded from rat orbitofrontal cortex (OFC) during an odor classification task~\citep{KepecsEtAl2008,KepecsOFC,KobakEtAl2016}.
The second data example explores the application of kdPCA to neurons recorded in macaque lateral intraparietal (LIP) cortex during a motion direction discrimination task~\citep{MeisterHennigHuk2013}.

\section{Methods}

\subsection{A summary of dPCA}

This section briefly describes dPCA from Kobak {\emph{et al}}.
The responses of $N$ neurons are placed in a matrix of observations, $\mx$, of size $M \times N$,
where $M$ is total the number of observations over all conditions.
The observations can be trial-averaged firing rates or single-trial responses of simultaneously recorded neurons.
For simplicity, I assume that the mean response of each neuron has been subtracted (i.e., the columns of $\mx$ have mean zero).
Each row of $\mx$ can be indexed by different discretely valued experimental parameters.
For brevity of notation, I consider only two parameters in this section: time ($t$) and stimulus ($s$).
However, this definition can be extended to include additional indices such as trial and decision as given in Kobak~\emph{et al}.
Let the row vector $\mx_{(t,s)} \in \mathbb{R}^N$ be the observed activity at time $t$ for stimulus $s$.
The observations are then be decomposed as the sum of averages over conditions plus an interaction term
\begin{align}
\mx_{(t,s)} = \langle \mx_{(t,s)} \rangle_t + \langle \mx_{(t,s)} \rangle_s + \epsilon_{t,s}
\end{align}
where the interaction term is $\epsilon_{t,s} = \mx_{(t,s)} - \langle \mx_{(t,s)} \rangle_t - \langle \mx_{(t,s)} \rangle_s$.

The next step is to construct the marginal matrices (each of size $M \times N$) over stimulus ($\mx_S$), time ($\mx_T$), and stimulus-time interaction ($\mx_{ST}$) by replacing the rows of $\mx$ with the averaged terms
\begin{align}
\mx_{S,(t,s)} & =  \langle \mx_{(t,s)} \rangle_t,\\
\mx_{T,(t,s)} & = \langle \mx_{(t,s)} \rangle_s, \nonumber\\
\mx_{ST,(t,s)} & = \epsilon_{t,s}. \nonumber
\end{align}

The goal of dPCA is to find a demixed low-rank reconstruction of $\mx$ by obtaining low-rank reconstructions of each marginal matrix.
For a choice of $R_\gamma \ll N$ for each $\gamma \in \{t,s,st\}$, dPCA fits decoding matrices $\md_\gamma$ and encoding matrices $\mf_\gamma$ of size $N\times R_\gamma$  that minimize the quantity
\begin{equation}
L_{dPCA} = \sum_{\gamma \in \{t,s,st\}} || \mx_\gamma - \mx\md_\gamma\mf_\gamma^\top||^2  + \mu ||\md_\gamma\mf^\top_\gamma||^2
\label{eq:dPCApenalty}
\end{equation}
where and $||\cdot||$ is the Frobenius norm.
The columns of $\mf_\gamma$ are constrained to be orthonormal.
The regularization term, $\mu = \frac{\lambda}{M} ||\mx||^2$, discourages overfitting and the variable $\lambda$ controls the strength of regularization (Kobak \emph{et al.} instead used $\mu = (\lambda ||\mx||)^2$).

Solving for $\mf_\gamma$  is accomplished by substituting $\md_\gamma\mf_\gamma^\top$ with the full-rank matrix $\mc_\gamma$ and then taking the standard regularized least-squares solution:
\begin{align}
\mc_\gamma = (\mx^\top\mx + \lambda\eye_N)^{-1}\mx^\top\mx_\gamma
\end{align}
 where $\eye_N$ is the $N\times N$ identity matrix.
The dPCA algorithm provided by Kobak \emph{et al.} completes the reduced-rank regression by performing PCA on $\mx\mc_\gamma$, taking the first $R$ principal components ($\mV_R$), and setting
\begin{align}
\md_\gamma & = \mc_\gamma\mV_R, & \mf_\gamma^\top & = \mV_R^\top
\end{align}
to recover the demixed principal components (dPCs).
The procedure is repeated for each $\gamma$.

\subsection{Dimensionality reduction with kdPCA}

The formulation of kdPCA presented in this section performs nonlinear dimensionality reduction to decode high-dimensional neural activity into the low-dimensional, demixed latent space.
To build upon dPCA, each observation (the activity of $N$ neurons in one time bin) is mapped into a new space by the function $\Psi: \mathbb{R}^N \rightarrow \mathcal{A}$ where $\mathcal{A}$ is a Hilbert space whose dimensionality ($\textrm{dim}(\mathcal{A}) = N^*$) may be greater than $N$.
Let the term $\Psi(\mx)$ denote the $M\times N^*$ matrix obtained by passing each row of $\mx$ through $\Psi$.
This new matrix is then used to reconstruct the marginal decomposition of $X$.
That is, kdPCA finds linear operators $\mh_\gamma$ that project the terms of $\Psi(X)$ into a low-dimensional space ($\mathbb{R}^{R_\gamma}$) and matrices $\mg_\gamma$ that reconstruct the parameter-dependent observations, $\mx_\gamma$.

To motivate how projecting data into a higher dimensional space may aid in demixing subspaces of neural activity with a concrete example, \figref[ left]{fig:kernelDemo} shows the activity of two hypothetical neurons ($x$ and $y$) over several trials in two stimulus categories (blue and red).
In the neural activity space, $\mathbb{R}^2$, the two stimulus categories cannot be linearly separated: no straight line can divide the red and blue points and standard PCA cannot find a subspace that separates the two classes.
Let $\Psi_{ex}: \mathbb{R}^2 \rightarrow \mathbb{R}^3$  be a polynomial expansion of the original space such that
\begin{align}
\Psi_{ex}(x,y) & = (x,y,x^2 + y^2).
\end{align}

In this transformed space, \figref[B]{fig:kernelDemo} shows that the conditions can now be separated linearly.
Additionally, performing PCA in this higher dimensional space reveals a single component that separates the red and blue classes~\citep{ScholkopfSmolaMuller1998}.
Thus, taking nonlinear functions of the neural activity can reveal components related to different experimental conditions that cannot be captured by purely linear methods.

\begin{figure}[t!]
\centering
\includegraphics[width=0.5\linewidth]{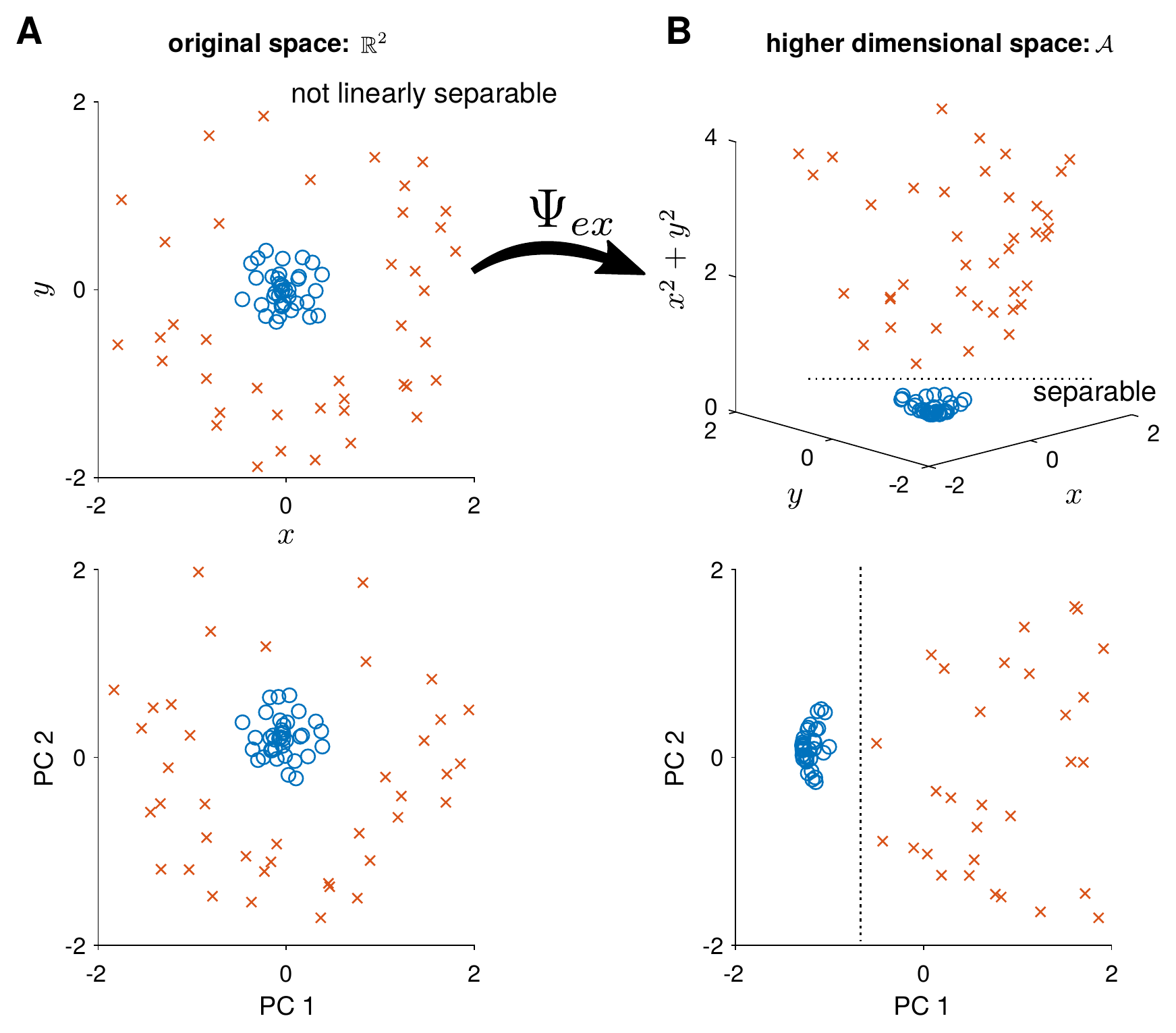}
\caption{A demonstration of kernel PCA.
{\bf A} (top) A two-dimensional space of neural activity.
Each color represents two possible conditions.
In this space, the red and blue conditions cannot be linearly separated.
Performing PCA (bottom) in the original space returns essentially the same representation.
This situation could potentially occur in a neural population exhibiting rotational dynamics with a different gain across conditions.
{\bf B} Projecting these observations into a higher dimensional space (top) through the function $\Psi$ allows these classes to be linearly separable.
In this higher dimensional space, the two classes are separated in the first principal component (bottom).
}
\label{fig:kernelDemo}
\end{figure}

Plugging $\Psi$ into the dPCA loss function produces the analogous loss function for kdPCA:
\begin{equation}
L_{kdPCA} = \sum_{\gamma \in \{t,s,st\}} || \mx_\gamma - \Psi(\mx)\mg_\gamma\mh_t^\top||^2 + \eta ||\mg_\gamma\mh_\gamma^\top||^2.
\label{eq:kdPCA}
\end{equation}
Similarly to $\md_\gamma$ in dPCA,  the decoding matrix $\mg_\gamma$ is a matrix of size $N^* \times R_\gamma$.
The encoding matrix, $\mh_\gamma$, is again of size $N \times R_\gamma$, and the columns are constrained to be orthonormal.
The solution is regularized with the penalty $\eta$.

The following observation is used to minimize the loss function: each column of $\mg$ must be in the row space of $\Psi(\mx)$.
Otherwise, the solution would include components outside the subspace of $\mathcal{A}$ explored in the data, and regularization will shrink weights in those directions to zero.
Therefore, for some matrix $\mz_\gamma$ of size $M \times R_\gamma$  the decoding matrix can be written as
\begin{equation}
\mg_\gamma = \Psi(\mx)^\top \mz_\gamma.
\label{eq:dual}
\end{equation}
Combining \eqsref{eq:kdPCA} and \ref{eq:dual} gives
\begin{align}
L_{kdPCA} & = \sum_{\gamma \in \{t,s,st\}} || \mx_\gamma - \Psi(\mx)\Psi(\mx)^\top \mz_\gamma\mh_t^\top||^2 + \eta || \Psi(\mx)^\top \mz_\gamma \mh_\gamma^\top||^2.
\label{eq:kdPCA2}
\end{align}

The elements of the matrix $\mk = \Psi(\mx)\Psi(\mx)^\top$ are dot products of rows of $\Psi(\mx)$, which can be computed using a kernel function $\kappa$ on the rows of $\mx$:
\begin{equation}
\mk_{i,j} = (\Psi(\mx)\Psi(\mx)^\top)_{i,j} = (\Psi(\mx)_{i,\cdot} \cdot \Psi(\mx)_{j,\cdot}) = \kappa(\mx_{i,\cdot},\mx_{j,\cdot})
\end{equation}
where $\mx_{j,\cdot}$ is the $j$th row of $\mx$.
This results in the loss function
\begin{align}
L_{kdPCA} & = \sum_{\gamma \in \{t,s,st\}} || \mx_\gamma - \mk \mz_\gamma\mh_\gamma^\top||^2 + \eta ||\Psi(\mx)^\top\mz_\gamma\mh_\gamma^\top||^2 \nonumber\\
 & = \sum_{\gamma \in \{t,s,st\}} || \mx_\gamma - \mk \mz_\gamma\mh_\gamma^\top||^2 + \eta \cdot \textrm{Trace}\left( \Psi(\mx)^\top\mz_\gamma\mh_\gamma^\top (\Psi(\mx)^\top \mz_\gamma\mh_\gamma^\top)^\top  \right) \nonumber\\
 & = \sum_{\gamma \in \{t,s,st\}} || \mx_\gamma - \mk \mz_\gamma\mh_\gamma^\top||^2 + \eta \cdot \textrm{Trace}\left( \mk\mz_\gamma\mh_\gamma^\top (\mz_\gamma\mh_\gamma^\top)^\top  \right).  
\label{eq:kdPCA3}
\end{align}

To solve for $\mh_\gamma$ and $\mz_\gamma$ (and implicitly $\mg_\gamma$), the same low-rank regression procedure used for dPCA is applied.
First, we consider the full-rank regularized least-squares solution by substituting $\mz_\gamma\mh_\gamma^\top$ with $\mc^*_\gamma$.
Differentiating \eqref{eq:kdPCA3} with respect to $\mc^*_\gamma$ is then
\begin{align}
\frac{d}{d\mc^*_\gamma} L_{kdPCA} & = -2\mk\mx^\top + 2\mk\mk\mc^*_\gamma  +  2\eta\mk\mc^*_\gamma.
\end{align}
By setting the derivative to zero and solving for $\mc^*_\gamma$ the solution becomes~\citep{FriedmanEtAl2001}
\begin{align}
\mc^*_\gamma  & = (\mk + \eta\eye_M)^{-1}\mx_\gamma.
\end{align}
Note that this differs slightly from the linear regularized least-squares solution.
This solution reveals that one does not need to represent explicitly elements of $\mathcal{A}$ or the function $\Psi$, and instead only the choice of the an appropriate kernel function is needed.
Reduced-rank regression is then completed for each $\gamma$ by performing PCA on $\mk\mc_\gamma^*$, taking the first $R$ principal components ($\mV_R$), and setting
\begin{align}
\mz_\gamma & = \mc^*_\gamma\mV_R, & \mh_\gamma & = \mV_R^\top.
\end{align}

A new observation, $\vx^*$, can be projected into each marginal low-dimensional space by applying both \eqref{eq:dual} and the kernel representation to get
\begin{align}
\Psi(\vx^*)\mg_\gamma & = \Psi(\vx^*)\Psi(\mx)^\top \mz_\gamma \nonumber \\
& = \mathbf{k}^* \mz_\gamma
\end{align}
where $\mathbf{k}^*$ is a row vector of length $M$.
Each element of $\mathbf{k}^*$ can be computed using the kernel between the new observation and the training observations:
\begin{equation}
\mathbf{k}^*_i = \kappa(\vx^*,\mx_{i,\cdot}).
\label{eq:testKDPCA}
\end{equation}

For the examples presented here, I consider linear and Gaussian kernels
\begin{align}
\kappa_{lin}(\vx,\vy) & = \vx\cdot\vy\\
\kappa_{Gauss}(\vx,\vy) & = \exp\left(-\frac{1}{2l^2} ((\vx-\vy)\cdot(\vx-\vy))\right) \nonumber
\end{align}
where $l$ is a length-scale parameter.
For consistency with dPCA, the penalty term is parameterized as a function of $\lambda$ by $\eta  = \lambda \frac{1}{M} \textrm{Trace}(\mk)$.
Here I assume the same kernel is applied to find the kernel demixed principal components (kdPCs) for all parameters (time, stimulus, and interaction).
I selected the Gaussian kernel so that $\mk$ will resemble the graph Laplacian used in the Laplacian eigenmap technique for finding low-dimensional manifolds embedded nonlinearly in a high-dimensional space~\citep{BelkinNiyogi2003}.
However, the user is free to select other kernel functions depending on what is appropriate for different applications~\citep{CristianiniEtAl2000}.

\subsection{Simulations}

I present simulated populations of $N=50$ neurons with responses that lie in a latent space of dimension $D$.
Each simulation consists of $S$ stimulus conditions, and each condition contains $T$ time points (observations).
The latent trajectory for condition $i$ is denoted $\ml_{i}$ which is of size $T \times D$.
The trajectories are approximately smooth in time in the examples explored here.
For the first three simulations, the dimensionality is $D=2$ while the fourth simulation uses a higher dimensional space ($D=6$).
The latent trajectories over all conditions are concatenated as the $M\times D$ matrix
\begin{equation}
 \ml = \left[\ml_1^\top, \ml_2^\top, \dots \ml_S^\top \right]^\top,
\end{equation}
where $M = T\cdot S$.
The contents of $\ml$ for each type of simulation are defined in the corresponding section.
 
The loading weights which linearly map the latent space $\ml$ to neural space is the $D \times N$ dimensional matrix $\mw$.
The elements of $\mw$ are drawn as independent standard normal variables.
The observed neural responses are generated by taking
\begin{align}
\mx^* & = \ml\mw + \sigma\mx_{noise},\\
{\mx_{noise}}_{(i,j)} & \displaystyle \underset{\scriptscriptstyle iid}{\sim} \mathcal{N}(0,1) \nonumber
\end{align}
where $\mx_{noise}$ is a $M\times N$ noise term drawn for each simulation.
The noise standard deviation was $\sigma = 1$.
Finally, the columns of $\mx^*$ were z-scored to produce the observation matrix, $\mx$.
In each type of simulation, $\ml$ was kept constant while the loading weights and noise were resampled independently for each repeat of the simulation.

For simplicity of presentation, all dPCA and kdPCA fits to the simulated examples used a regularization parameter of $\lambda = 1$.
The length scale of the Gaussian parameter was fixed to $l=5$.

The code to generate the simulations and perform all analyses is available publicly at \url{https://github.com/latimerk/kdpca}.

\subsubsection{Assessing demixing performance in simulations}

I quantified the neural activity projected onto the first time component and the first stimulus components using two metrics to compare the interpretability and demixing performance dPCA and kdPCA on the simulated cases.
These particular measures are not intended to extend to real data applications. 

The performance for time was measured by the $R^2$ computed by linear regression between time and the neural activity projected onto the first time component.
This measures both how well neural activity is mapped into a space that depends on time, and how similarly all the stimulus conditions are mapped into this space.
The test performance was computed as the $R^2$ between the test conditions projected onto the first component and the least-squares regression line computed on the training conditions as a function of time.

To assess how well the stimulus dimension separates stimulus conditions, I computed the minimum $d'$~\citep[sensitivity index;][]{MacmillanEtAl2004} of the observations projected onto the first stimulus dimension ($\md_s$ and $\mg_s$) score between each stimulus condition:
\begin{align}
d'_{s,dPCA} & = \min_{i\neq j} |d'(\mx_{(\cdot,i)}\md_s,\mx_{(\cdot,j)}\md_s )|,\\
d'_{s,kdPCA} & = \min_{i\neq j}|d'(\Psi(\mx_{(\cdot,i)})\mg_s,\Psi(\mx_{(\cdot,j)})\mg_s )| \nonumber
\end{align}
where $\mx_{(i,\cdot)}$ is the set of all observations of stimulus condition $i$.
The sensitivity index is computed as the difference in means between two conditions normalized by the variance within conditions:
\begin{align}
d'(\mathbf{a},\mathbf{b}) & = \frac{\langle\mathbf{a}\rangle-\langle\mathbf{b}\rangle}{\sqrt{\frac{1}{2}\left(\textrm{Var}(\mathbf{a}) - \textrm{Var}(\mathbf{b}) \right)}}.
\end{align}

This metric measures the dependency of time in the stimulus component (i.e., the variance) relative to the mean separation between two stimulus conditions, and takes the worst pairwise case over all conditions.
Performance was measured on the training conditions alone (training), and by testing how well the test conditions were separated from all training and test conditions (test).

\subsubsection{Variance explained by demixed components}

The percent of variance explained for an individual component in dPCA or kdPCA was computed as
\begin{align}
{VE}^{dPCA}_{(\gamma,j)} & = 100\times\left(1- \frac{|| \mx - \mx\md_{\gamma,j}\mf_{\gamma,j}^\top||^2}{||\mx||^2}\right)\\
{VE}^{kdPCA}_{(\gamma,j)} & = 100\times\left(1 - \frac{|| \mx - \mk\mz_{\gamma,j}\mh_{\gamma,j}^\top||^2}{||\mx||^2}\right) \nonumber
%{VE}^{dPCA}_{(\gamma,j)} & = 100\times\left(1- \frac{|| \mx - \mx\md_{\gamma,j}\mf_{\gamma,j}^\top||^2}{||\mx - \mathbf{1}_{M}\left(\frac{1}{M} \mathbf{1}_M^\top \mx\right)||^2}\right)\\
%{VE}^{kdPCA}_{(\gamma,j)} & = 100\times\left(1 - \frac{|| \mx - \mk\mz_{\gamma,j}\mh_{\gamma,j}^\top||^2}{||\mx - \mathbf{1}_{M}\left(\frac{1}{M} \mathbf{1}_M^\top \mx\right) ||^2}\right)
\end{align}
where $\md_{\gamma,j}$ and $\mf_{\gamma,j}$ (or $\mz_{\gamma,j}$ and $\mh_{\gamma,j}$) are the $j$th columns of the decoder and encoder matrices for parameter $\gamma$.
To compute the variance explained in the test conditions for kdPCA, the $\mk$ term was replaced by $\mathbf{k}$ given in \eqref{eq:testKDPCA}.
This formula assumes that the columns of $\mx$ had zero mean.

\subsection{Neural data}
\subsubsection{Orbitofrontal cortex recordings}

I analyzed the trial-averaged response of 214 OFC neurons recorded from three rats while the animals performed an odor discrimination task.
These data have been reported previously and were obtained from \url{http://crcns.org}~\citep{KepecsEtAl2008,KepecsOFC}.
This data set was one of the four examples used by Kobak {\emph{et al.}} to demonstrate dPCA.
The data were processed as described by Kobak \emph{et al.} using code made available at \url{https://github.com/machenslab/elife2016dpca}.

The length-scale of the Gaussian kernel was set to $l=50$ and the analyses are reported over a range of regularization parameters settings.

\subsubsection{Lateral intraparietal cortex recordings}

The second data example considered the responses of LIP neurons recorded from two monkeys while the animals performed a random dot motion discrimination task.
These data were reported in \cite{MeisterHennigHuk2013}.
The neurons were recorded individually, and I therefore used trial-averaged firing rates across each condition. 
The firing rates were smoothed by a Gaussian filter with a width of $50$~ms.
Cells were included in the analysis if at least one trial was recorded for every condition considered, and 69 out of 80 cells met this criterion.
The data are available alongside the kdPCA code at \url{https://github.com/latimerk/kdpca}.

For kdPCA, the length-scale parameter of the Gaussian kernel was again set to $l=50$ and the regularization parameter was set to $\lambda = 1.0$ (however, the results presented here did not strongly depend on the choice of $\lambda$). 
The dPCs are shown over a large range of $\lambda$, rather than fitting with a single selection of $\lambda$.

I examined whether dPCs across different experimental parameters were orthogonal to determine the independence of encoding axes.
Orthogonality is measured as the dot product between the first dPC for parameter $a$ (denoted $\mf_{1,a}$; the first column of the encoding matrix) and the first dPC for parameter $b$. 
Following Kobak {\emph{et al.}}, the angle between encoding vectors was deemed significantly non-orthogonal at $p<0.001$ if $|\mf_{1,a}\cdot\mf_{1,b}|>3.3/\sqrt{N}$ (note that the columns of $\mf_{\gamma}$ are unit vectors).

\subsection{Regularization and cross-validation}

Both dPCA and kdPCA require selecting the penalty term ($\lambda$) and number of components ($R_\gamma$).
Additionally, kdPCA requires selecting any kernel parameters such as the length scale.
One powerful approach for selecting such terms is cross-validation~\citep{JamesEtAl2013}.
Although a complete treatment of cross-validation for demixed dimensionality reduction is beyond the scope of this paper, this section provides a brief outline of cross-validation for dPCA or kdPCA along with references for users to consider when applying either method to data.
 
Kobak {\emph{et al.}} proposed a cross-validation technique for selecting the penalty term that leveraged the fact that elements of $\mx$ are often trial-averaged responses, especially when analyzing neurons that were not recorded simultaneously.
 Leave-one-out CV sets are then formed by taking one trial for each condition from each neuron to form $\mx_{test}$.
 The remaining trials are then averaged to create complete $\mx_{train}$.
 Thus, the test and train matrices have the same elements and the dPCA reconstruction of $\mx_{test}$ can be treated as the estimate of $\mx_{train}$.
 Kobak {\emph{et al.}} selected $\lambda$ to minimize the cross-validation score
\begin{align}
L_{cv}(\lambda) = \frac{\sum_{\gamma} || \mx_{\lambda,train} - \mx_{test}\md_\gamma(\lambda)\mf_\gamma(\lambda)||^2}{||\mx_{train}||^2}
\end{align}
where $\md_\gamma(\lambda)$ and $\mf_\gamma(\lambda)$ are decoder and encoder matrices trained with the regularization parameter $\lambda$.
Because this cross-validation score aims to predict the marginal matrices, the score considers both the demixing performance and the total reconstruction error.
The user can select a single $\lambda$ value or select distinct $\lambda$ for each parameter $\gamma$.
 
The approach given by Kobak {\emph{et al.}} differs from cross-validation methods applied more generally to PCA-style dimensionality reduction where the observations cannot be assumed to be averages over multiple trials.
Care must be taken to ensure that training and test sets are fully separated in such cases~\citep{Wold1978,OwenPerry2009}.
\cite{BroEtAl2008} provides a review of cross-validation techniques for selecting the rank in PCA based on leaving out entire observations (i.e., completely leave out an element of the data matrix $\mx$).
In these methods, a low-rank decomposition of the data matrix is formed without considering the left-out elements.
The left-out elements are then estimated (filled-in) using the remaining elements that share the same row or column of $\mx$.
Importantly, complete rows or columns of $\mx$ cannot be left out without fitting new parameters to  the test data: a violation of cross-validation.
Similar leave-out methods have been proposed for selecting penalty terms in penalized matrix decomposition~\citep[PMD;][]{WittenEtAl2009} and penalized tensor decomposition~\citep{MadridScott2017}.

Applying established cross-validation approaches from PCA and PMD requires extra care for (k)dPCA, because the marginal matrices, $\mx_\gamma$, are computed by averaging across conditions.
Thus, one cannot simply leave out an element from each marginal matrix for cross-validation because elements across and within marginal matrices are dependent.
Furthermore, leaving out elements of the data matrix, $\mx$, and then recomputing the marginal matrices could bias the marginalization procedure due to the missing data.
However, model-based approaches for computing the marginal matrices in the presence of missing data (such as the linear model method given in Kobak {\emph{et al.}}) could be applied to the partial data matrix for each fold.
Model-based approaches to handling missing data are therefore one potential route for extending cross-validation techniques from PCA and PMD to the penalized (k)dPCA case.
These cross-validation techniques should be considered more thoroughly in future work on demixed dimensionality reduction, particularly when considering single-trial observations of simultaneously recorded neurons~\citep{WilliamsEtAl2018}.
Additionally, the evaluation of cross-validation performance should take into account both demixing performance (e.g, assess whether the time components independent of stimulus) and reconstruction error.

\section{Results}

\subsection{Example 1: Low-dimensional summation of components}

The first simulated example demonstrates a simple case with linearly demixable stimulus and time components (\figref[A]{fig:linearSim}).
Each 15 time-point trajectory follows a simple linear path which is translated in a nearly orthogonal dimension by the stimulus condition.
This scenario is the optimal scenario for dPCA: the latent space can be decomposed linearly into a single time plus a single stimulus component.
The activity of 50 simulated neurons depends linearly on the points in this subspace plus Gaussian noise.

\begin{figure}[t]
\centering
\includegraphics[width=0.8\linewidth]{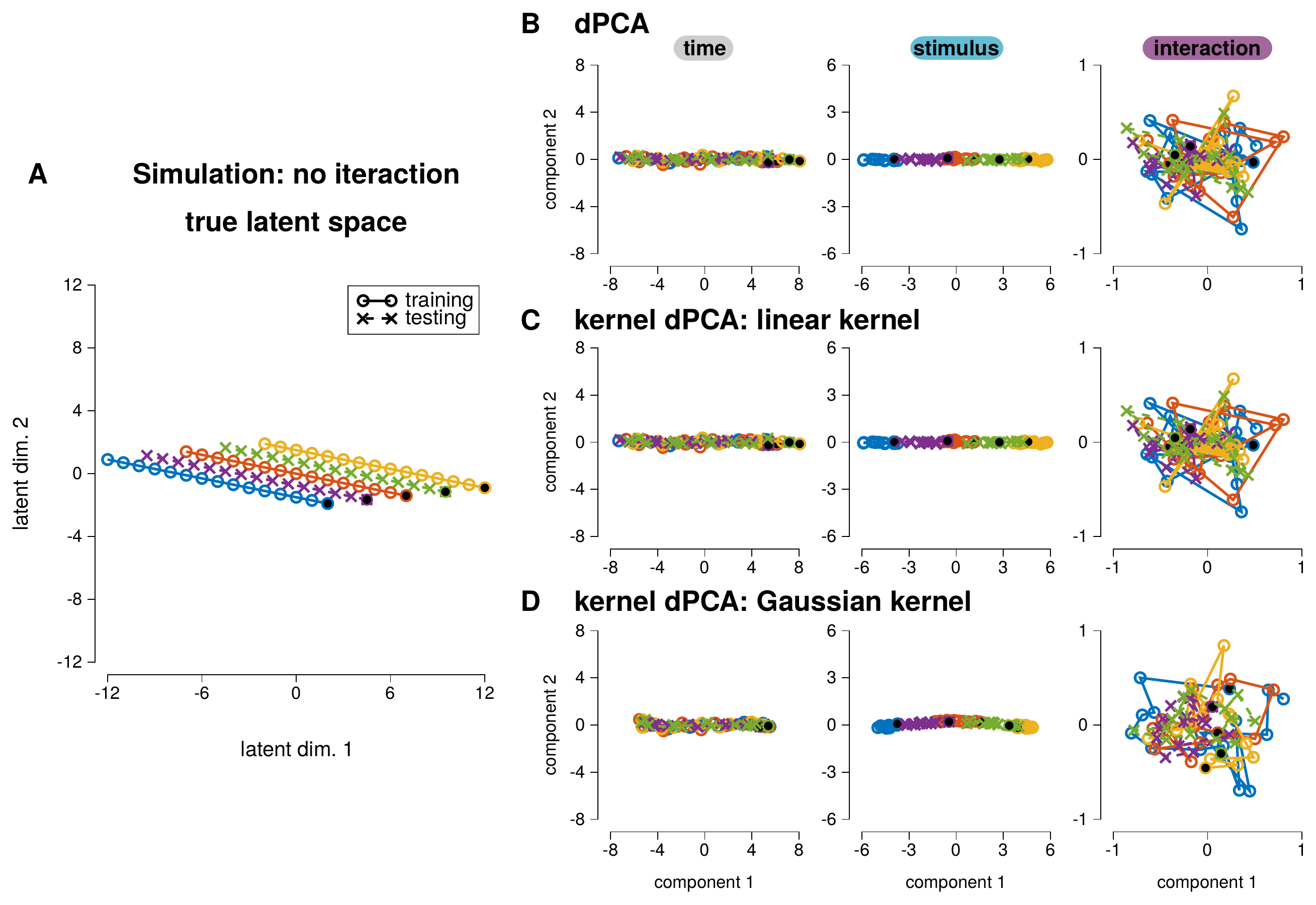}
\caption{Two-dimensional subspace representing latent neural activity as the sum between stimulus condition (indicated by color) and time.
({\bf A}) The true subspace used to generate the simulated neural activity.
The circles show trajectories used for finding the low dimensional space, and the crosses indicate test conditions only used to evaluate the estimated components.
The black marks denote the time point in each trajectory.
({\bf B}) The simulated activity projected onto the first two dPCs for time (left), stimulus (middle), and interaction (right). 
({\bf C}) Same as B for the kdPCs with a linear kernel.
The identical results of B and C demonstrate that linear kdPCA is in fact equivalent to dPCA.
({\bf D}) Same as C for the kdPCs with a Gaussian kernel.
}
\label{fig:linearSim}
\end{figure}

Here, I compare dPCA to kdPCA with both a linear and a Gaussian kernel on the simulated activity.
Only the three training trajectories (circles) were used to fit dPCA and kdPCA.
The activity of the neurons projected onto the first two dPCs and kdPCs are shown in \figref[B-D]{fig:linearSim}.
The results demonstrate the equivalence of dPCA and kdPCA with a linear kernel (in this case, $\Psi$ is the identity function).
Thus, the kdPCA family contains dPCA as a special case.
The time and stimulus components from both dPCA and kdPCA fall primarily along the first dimension.
The interaction dimensions in all three methods captures noise in the population activity.
The percent of variance of the population activity explained by the 1st stimulus, time, and interaction dimension was nearly identical for all methods (\figref[A,D,G]{fig:linearSimComponents}; \tableref{tab:perVarExplained}).

\begin{figure}[t!]
\centering
\includegraphics[width=0.9\linewidth]{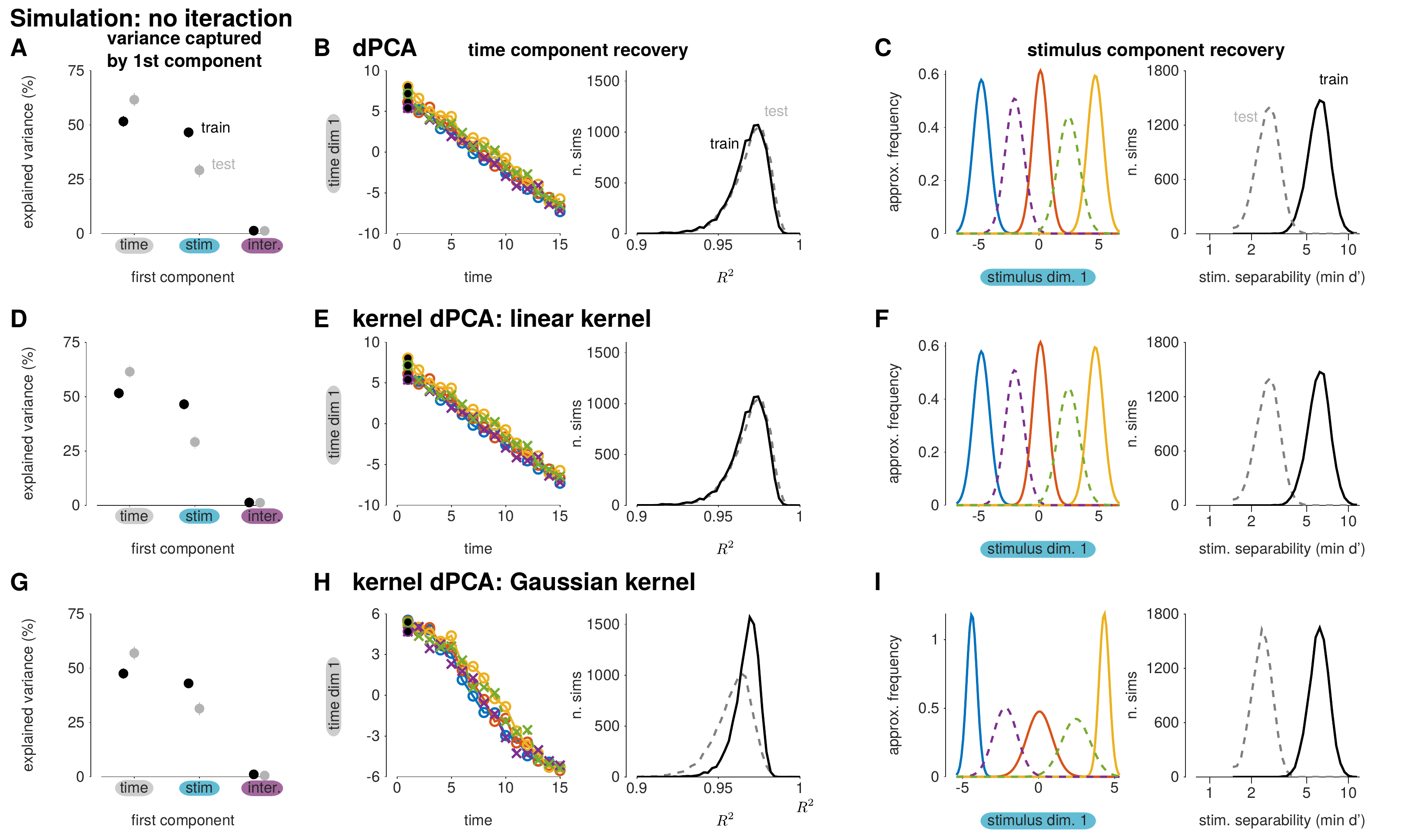}
\caption{ Recovery of time and stimulus with the first components from dPCA and kdPCA for the linearly demixable simulation in \figref{fig:linearSim}.
({\bf A}) The mean percent variance explained in the neural activity by the first dPCA time, stimulus, and interaction components.
The black points show the percent explained in the training data and the test data are shown in gray.
Error bars indicate the 50th percentile region over the 10000 simulations.
({\bf B}) (left) For the example simulation in \figref{fig:linearSim}, the recovered latent trajectory of first time dPC as a function of time.
(right) Distribution of the $R^2$ of the first time component with time (squared Pearson's correlation of the two axes in A) for the 10000 simulated neural populations.
Solid distribution shows the correlations calculated on the training conditions
and the dash distribution shows the correlations for the test conditions.
({\bf C}) (left) For the example simulation in \figref{fig:linearSim}, the distribution of neural activity projected onto the first stimulus dPC for each stimulus condition approximated by a Gaussian distribution.
(right) Quantifying the separability of the stimulus as the minimum d' between each stimulus in the first stimulus dPC over 10000 simulations.
Solid distribution shows the separability calculated on the training conditions
and the dash distribution shows the separability for the test conditions.
({\bf D-F}) Same as A-C for the first linear kdPC.
({\bf G-I}) Same as A-C for the first Gaussian kdPC.
}
\label{fig:linearSimComponents}
\end{figure}

To quantify the demixing of the temporal and stimulus components, I examined the activity projected onto the first time component.
For the simulation in \figref[A]{fig:linearSim}, time and the first dPC are well correlated for all stimuli (\figref[B,E,H left]{fig:linearSimComponents}).
The $R^2$ across all stimulus conditions measures how the time component maps all stimuli onto a common path that depends linearly on time.
For 10000 simulated populations of 50 neurons, all three methods found a time component with strong correlation with time in the three training conditions \figref[B,E,H right]{fig:linearSimComponents},  \tableref{tab:demixingPerformance}).
These results were consistent when predicting the test stimulus time trajectories (denoted by crosses).

I then examined how the first stimulus dimension separated the stimulus conditions.
For each stimulus condition, I projected the 15 time points onto the first stimulus dimension, and I approximated each condition as a normal distribution on this dimension (\figref[C,F,I left]{fig:linearSimComponents}).
If two distributions show little overlap, the two stimulus conditions are separated in the 1st stimulus component across all time points.
In contrast, overlapping distributions would show that the component does not demix time from stimulus successfully.
Each method was able to separate the three training stimulus conditions successfully (\figref[C,F,I right]{fig:linearSimComponents};   \tableref{tab:demixingPerformance}).
Moreover, the separability of stimulus category in the first component generalized to the test stimuli.

In summary, dPCA and kdPCA recovered the independent stimulus and time dimensions in this simulation.
kdPCA with a linear kernel produced identical results to dPCA, and thus provides an alternative formulation of dPCA that scales in the number of observations instead of the number of neurons.
This scaling could aid in finding linear components as the number of recorded neurons in a single experiment grows~\citep{StevensonKording2011,HainmuellerHazlett2014,RoduEtAl2018}.
Additionally, kdPCA with a Gaussian kernel achieved close to the same performance as dPCA (the optimal method for this particular example).

\begin{table}[t!]
  \begin{center}
    \begin{tabular}{l l  c r |  c r} 
                 &             & \multicolumn{2}{c }{time $R^2$}  &  \multicolumn{2}{c}{stimulus $d'$}\\ 
      Simulation & Algorithm   & train & \multicolumn{1}{c }{test}                    & train & \multicolumn{1}{c}{test} \\ \hline \hline
      Linear & dPCA           & $ 0.97 \pm 0.01$ & $ 0.97 \pm 0.01 $ & $6.22 \pm 1.14$ & $2.67 \pm 0.52$\\
             & linear kdPCA   & $ 0.97 \pm 0.01$ & $ 0.97 \pm 0.01 $ & $6.22 \pm 1.14$ & $2.67 \pm 0.52$\\
             & Gaussian kdPCA & $ 0.97 \pm 0.01$ & $ 0.96 \pm 0.01 $ & $6.21 \pm 1.03$ & $2.41 \pm 0.42$\\
      \hline
      Rotation & dPCA         & $0.09 \pm 0.10$ & $-0.26 \pm 0.34$ & $1.56 \pm 0.91$ & $0.51 \pm 0.40$\\
             & Gaussian kdPCA & $0.88 \pm 0.11$ & $ 0.48 \pm 0.36$ & $3.27 \pm 2.03$ & $2.03 \pm 0.92$\\
      \hline
      Scaling & dPCA           & $0.86 \pm 0.01$ & $0.93 \pm 0.01$ & $0.85 \pm 0.07$ & $0.38 \pm 0.04$\\
              & Gaussian kdPCA & $0.97 \pm 0.01$ & $0.97 \pm 0.01$ & $6.35 \pm 0.48$ & $2.81 \pm 0.26$\\
    \end{tabular}
    \caption{Summary of the demixing performance metrics of dPCA and kdPCA for the three simulations.
    Each value shows the mean and standard deviation of the metric over 10000 randomly generated simulations.}
    \label{tab:demixingPerformance}
  \end{center}
\end{table}

\begin{table}[h!]
  \begin{center}
    \begin{tabular}{l l  r r |  r r |  r r} 
                &             & \multicolumn{2}{c }{time} & \multicolumn{2}{c }{stimulus} & \multicolumn{2}{c}{interaction}\\ 
     Simulation & Algorithm   & \multicolumn{1}{c}{train}        & \multicolumn{1}{c  }{test}        & \multicolumn{1}{c}{train}         & \multicolumn{1}{c }{test}           & \multicolumn{1}{c}{train}        & \multicolumn{1}{c}{test} \\ \hline \hline
      Linear & dPCA           &$ 52 \pm 4 $ & $62 \pm 4 $  &$ 47 \pm 4 $ & $29 \pm 5 $  &$ 1 \pm 0 $ & $1 \pm 0 $  \\ 
            & linear kdPCA    &$ 52 \pm 4 $ & $62 \pm 4 $  &$ 47 \pm 4 $ & $29 \pm 5 $  &$ 1 \pm 0 $ & $1 \pm 0 $  \\ 
             & Gaussian kdPCA &$ 48 \pm 3 $ & $57 \pm 4 $  &$ 43 \pm 3 $ & $31 \pm 4 $  &$ 1 \pm 0 $ & $1 \pm 0 $  \\ 
      \hline
      Rotation & dPCA         &$ 1 \pm 0 $ & $1 \pm 0 $  &$ 44 \pm 3 $ & $40 \pm 28 $  &$ 12 \pm 1 $ & $11 \pm 8 $  \\ 
             & Gaussian kdPCA &$ 1 \pm 0 $ & $0 \pm 0 $  &$ 39 \pm 3 $ & $31 \pm 21 $  &$ 8 \pm 1 $ & $-1 \pm 2 $  \\ 
      \hline
      Scaling & dPCA           &$ 57 \pm 5 $ & $60 \pm 5 $  &$ 20 \pm 3 $ & $17 \pm 3 $  &$ 15 \pm 1 $ & $16 \pm 2 $  \\ 
              & Gaussian kdPCA &$ 49 \pm 4 $ & $58 \pm 4 $  &$ 13 \pm 2 $ & $7 \pm 2 $  &$ 8 \pm 1 $ & $3 \pm 1 $  
    \end{tabular}
   \caption{The percent variance explained by the first time, stimulus, and interaction dPC and kdPC for the three types of simulations.
    Each value shows the mean and standard deviation of the metric over 10000 randomly generated simulations.}
    \label{tab:perVarExplained}
  \end{center}
\end{table}

\subsection{Example 2: Rotations}

The second example shows a two-dimensional latent space with stimulus conditions that are rotated around the origin (\figref[A]{fig:rotatedSim}).
Each condition is a straight line, but unlike the first simulation example, the paths are no longer parallel.
This poses a potential challenge for finding interpretable subspaces that demix time and stimulus information, despite the fact that the true subspace contains clear structure and separability of components.

I applied dPCA and kdPCA (with a Gaussian kernel only) to 10000 simulations of 50 neurons that depended linearly on this subspace (\figref[B,D]{fig:rotatedSim}).
The first dPC and kdPC for each parameter explained on average a similar amount of variance in the training data (\figref[A]{fig:rotatedSimComponents}A; \tableref{tab:demixingPerformance}), but kdPCA explained less variance in the test conditions, indicating that some overfitting occurred (\tableref{tab:perVarExplained}).
However, the stimulus dPCs do not clearly demix the stimulus from time, and instead dPCA returns a close reconstruction of the original mixed subspace (\figref[B middle]{fig:rotatedSim}).
In contrast, the interaction components recovered by kdPCA appear to contain the rotational information: this can be visualized more clearly by summing the stimulus and interaction spaces (\figref[E]{fig:rotatedSim}).

\begin{figure}[t!]
\centering
\includegraphics[width=0.9\linewidth]{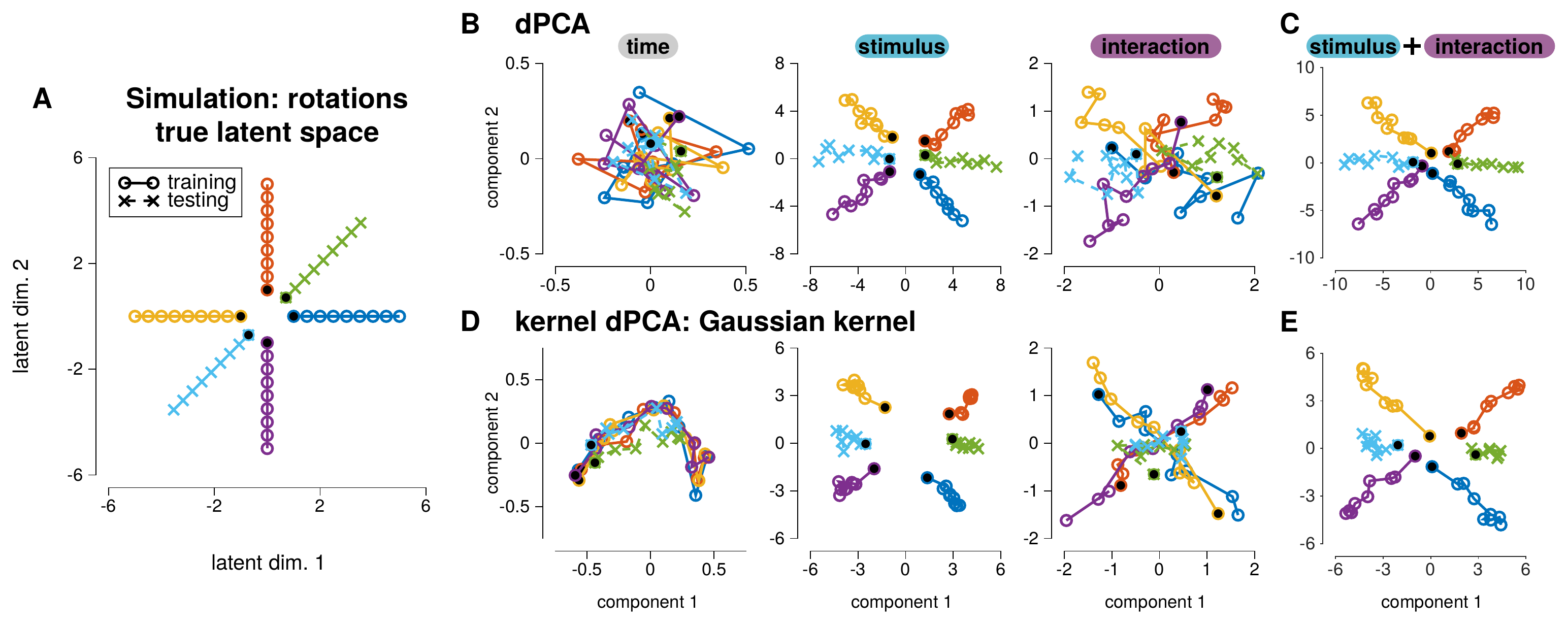}
\caption{Two-dimensional subspace representing latent neural activity for which the trajectories are rotated for each stimulus condition.
({\bf A}) The true subspace used to generate the simulated neural activity.
The circles show trajectories used for finding the low dimensional space, and the crosses indicate test conditions only used to evaluate the estimated components.
The black marks denote the time point in each trajectory.
({\bf B}) The simulated activity projected onto the first two dPCs for time (left), stimulus (middle), and interaction (right).
({\bf C}) The sum of the stimulus and interaction components (i.e., component 1 is the sum of the projection onto the first stimulus component and the projection onto the first interaction component).
({\bf D-E}) Same as B-C for the kdPCs with a Gaussian kernel.
}
\label{fig:rotatedSim}
\end{figure}

The demixed time components were again analyzed by computing the $R^2$ between time and the activity projected onto the first time component (\figref[B,E]{fig:rotatedSimComponents}).
Although the average of the four training conditions is a single point at the origin, the individual simulations contained noise.
Therefore, this analysis could still reveal how the radial structure of the simulation was reflected in the map to the noisy time average.
In this example, the time dPC failed to show a strong linear relationship to time in either the training or test conditions (\tableref{tab:demixingPerformance}).
However, the first time component recovered by kdPCA showed a strong relationship which generalized to the testing set.
To quantify the quality of the demixing in the stimulus subspace, I again took the minimum $d'$ across stimulus conditions on the first stimulus component (\figref[C-F]{fig:rotatedSimComponents}).
Gaussian kdPCA showed a higher degree of separation of stimulus conditions than dPCA.

\begin{figure}[t!]
\centering
\includegraphics[width=0.9\linewidth]{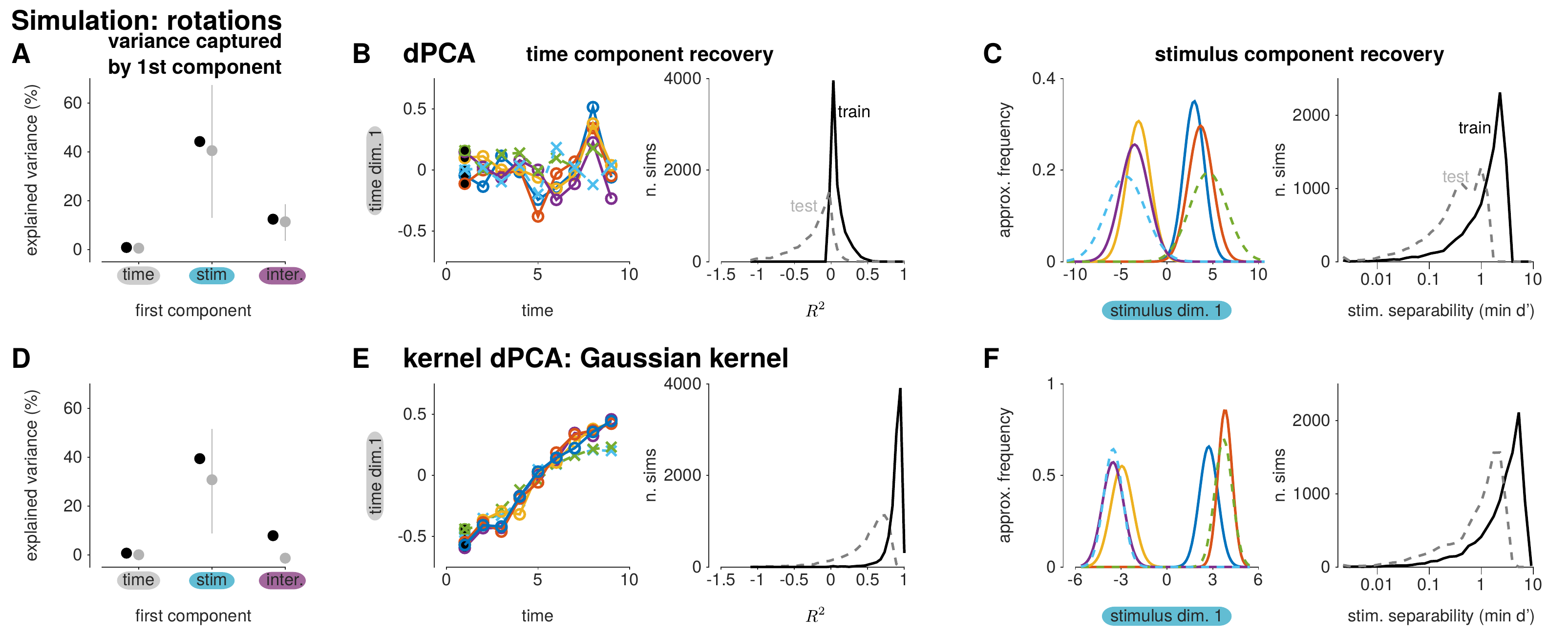}
\caption{ Recovery of time and stimulus with the first components from dPCA and kdPCA for the stimulus-dependent rotation simulation in \figref{fig:rotatedSim}.
({\bf A-C}) Analyses of the dPCA of the rotation simulation.
 Figures follow the structure as \figref[A-C]{fig:linearSimComponents} .
({\bf D-F}) Same as A-C for Gaussian kdPCA of the rotation simulation.
}
\label{fig:rotatedSimComponents}
\end{figure}

In this simulation, the demixed components discovered by dPCA differed quantitatively and qualitatively from the components derived by kdPCA with a Gaussian kernel.
kdPCA reliably discovered a nonlinear component that contains information about time in the simulated neural activity.
The stimulus dPCs recovered the complete structure of the original two-dimensional latent space.
In contrast, the interaction kdPCs visualized more clearly the condition-dependent rotated structure of the trajectories.

\subsection{Example 3: Scaling}

In this section, the two-dimensional latent space contained conditions that were scaled differently across stimulus conditions (\figref[A]{fig:scaledSim}).
Applying dPCA to simulated neural responses recovers time and interaction components that reflect the shape of the true latent 2-D space (\figref[B]{fig:scaledSim}).
This is expected from a linear method because both variables (stimulus and time) span that space.
However, the goal of demixed dimensionality reduction is to break this space up into relevant components and the dPCA result here acts more like standard PCA.
In contrast, kdPCA recovers a nonlinear embedding within the 2-D space that show greater demixing between time and stimulus (\figref[D]{fig:scaledSim}).
The interaction kdPCs reflect the scaling: the gold and blue traces --- the largest and smallest components respectively --- are the largest two traces in the interaction space.
In the 2-D kdPCA interaction space, the orientations of the gold and blue traces are flipped corresponding to the relative shrinking of the blue trace and stretching of the gold. 
The interaction component for the red condition, which is the mean of all the conditions, falls around the origin, consistent with no interaction for the average component.
Taking the sum of the interaction and time spaces shows that the interaction kdPCs indeed scale the time trace for each stimulus condition, resulting in a space that resembles the scaling in the true generating subspace (\figref[E]{fig:scaledSim}).

I again quantified the demixing performance of the first time and stimulus dimension of both techniques.
Each condition projected onto the first time dPC is well correlated with time individually, but the slope varies across stimulus conditions (\figref[B]{fig:scaledSimComponents};\tableref{tab:demixingPerformance}).
The kdPCA time component again shows a strong correlation with time, but the time component shows less dependence on the stimulus condition (\figref[B,E]{fig:scaledSimComponents}).
Separability of the stimulus conditions was also improved in the first stimulus kdPC compared to the first dPC (\figref[C,F]{fig:scaledSimComponents}).
The interaction and stimulus kdPCs also explained less variance in the data than dPCs (\figref[A,D]{fig:scaledSimComponents}; \tableref{tab:perVarExplained}).
This could occur if the variance explained by the stimulus and interaction dPCs included activity that could be accounted for by time alone.
Thus, kdPCA recovered subspaces that better captured the contributions of individual parameters and the scaling interaction than dPCA.

\begin{figure}[t!]
\centering
\includegraphics[width=0.9\linewidth]{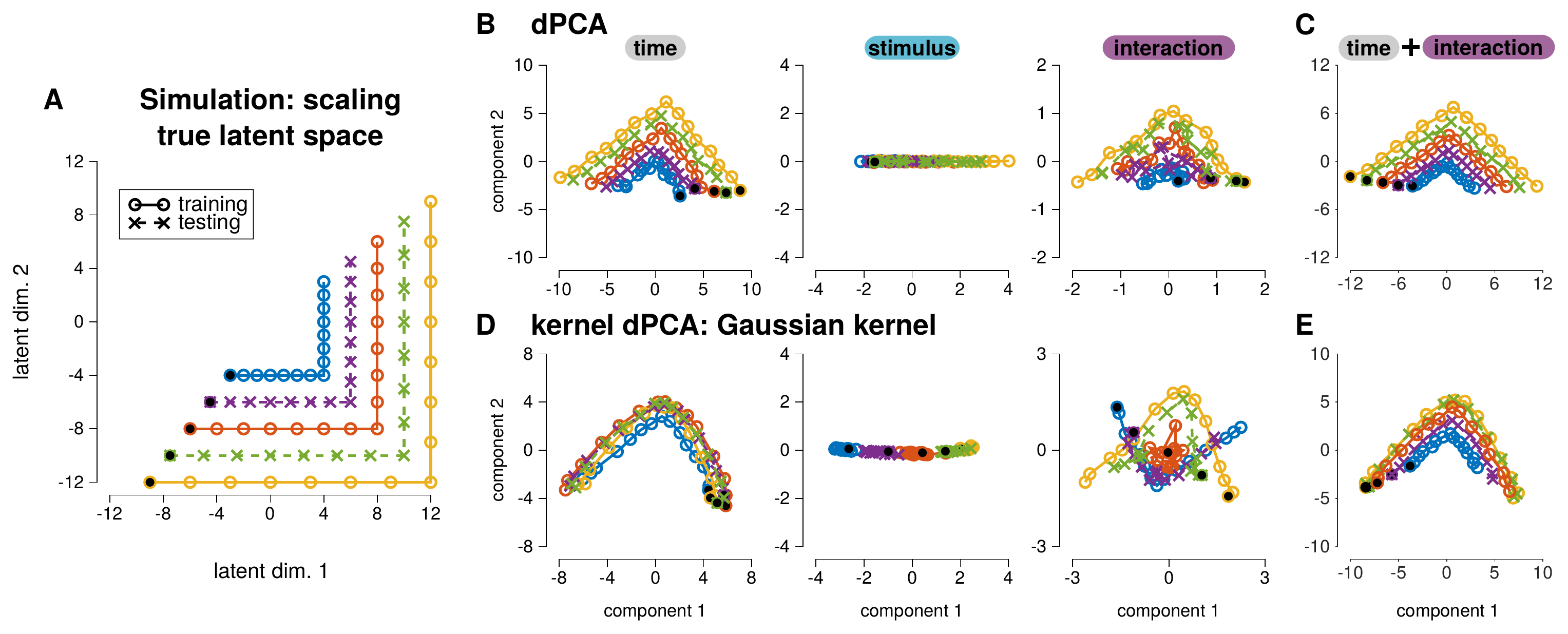}
\caption{Two dimensional subspace representing latent neural activity for which the trajectories are scaled for each stimulus condition.
({\bf A}) The true subspace used to generate the simulated neural activity.
The circles show trajectories used for finding the low dimensional space, and the crosses indicate test conditions only used to evaluate the estimated components.
The black marks denote the time point in each trajectory.
({\bf B}) The simulated activity projected onto the first two dPCs for time (left), stimulus (middle), and interaction (right). 
({\bf C}) The sum of the time and interaction components (i.e., component 1 is the sum of the projection onto the first time component and the projection onto the first interaction component).
({\bf D-E}) Same as B-C for the kdPCs with a Gaussian kernel.
}
\label{fig:scaledSim}
\end{figure}

\begin{figure}[ht]
\centering
\includegraphics[width=0.9\linewidth]{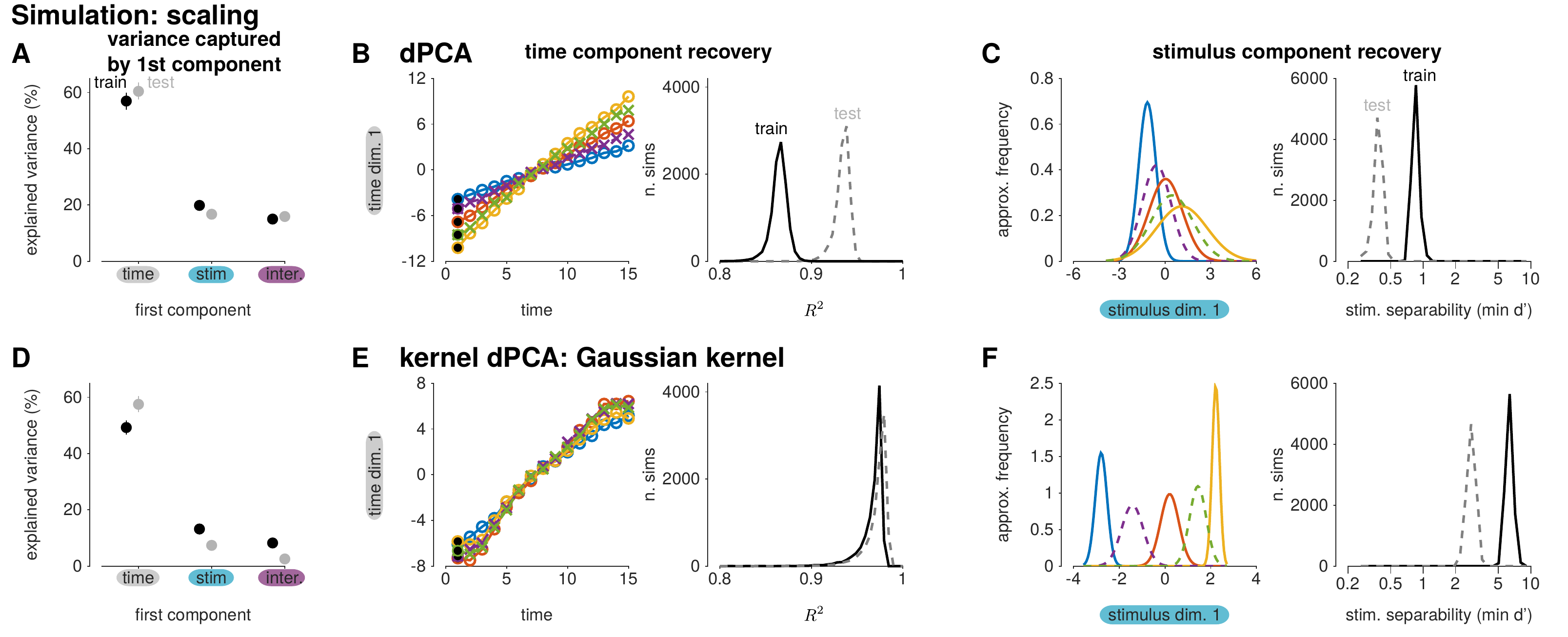}
\caption{ Recovery of time and stimulus with the first components from dPCA and kdPCA for the stimulus-dependent scaling simulation in \figref{fig:scaledSim}.
({\bf A-C}) Analyses of the dPCA of the scaling simulation.
 Figures follow the structure as \figref[A-C]{fig:linearSimComponents}.
({\bf D-F}) Same as A-C for Gaussian kdPCA of the scaling simulation.
}
\label{fig:scaledSimComponents}
\end{figure}

\subsection{Example 4: Scaling in higher dimensions}

Standard PCA could reveal the entire structure of the previous examples because the latent space was only two-dimensional for demonstration purposes.
Therefore, I extended the scaling simulation into a latent space of 6 dimensions to explore how kdPCA can help reveal structure in cases when data visualization is nontrivial.
For this simulation, the $d$th dimension of the latent space for conditions $s$ was defined as a function of time
\begin{align}
\ml_d(t) & = g(d,s)\cdot\left[\text{max}(0,\text{min}(10, t - 10\cdot(d-1)))) - 5\right] ,\\
g(d,s) & = 0.35s + 0.3d - 0.1ds - 0.05, \nonumber
\end{align}
for $t = 1 $ to $60$ and $s = 1$ to $5$.
This extends the L-shape of the 2-D scaling example (\figref[A]{fig:scaledSim}), because the latent trajectories move constantly in one dimension at a time for 10 time points starting from $d=1$ to $d=6$.
The function $g(d,s)$ scales each dimension by $0.5-1.5$ depending on the condition.
Importantly, the third condition is not scaled ($g(d,s=3) = 1$ for all $d$), and the scaling is symmetric around the third condition: that is $g(d,s=2) = g(d,s=4)$ and $g(d,s=1) = g(d,s=5)$ for all $d$.
Conditions 1, 3 and 5 were used to fit PCA, dPCA, and kdPCA, and conditions 2 and 4 were held as test conditions.

This example generates a 2-D manifold embedded nonlinearly into a 6-D space.
Other nonlinear dimensionality reduction methods such as t-SNE~\citep{MaatenHinton2008} could unravel the 2-D manifold in the 6-D space generated by this simulation.
These representations can be useful for visualization, even though these methods are not designed to return a demixed set of dimensions. 
However, the low-dimensional space discovered by t-SNE may not visually disambiguate the type of condition-dependent effects such as simple translation (which corresponds to no interaction term in dPCA) or scaling (which corresponds to interaction terms that share the shape of the time components).

Reducing the dimensionality using PCA on the simulated neural activity generated by these latent states shows condition-dependent activity across several dimensions (\figref[A-B]{fig:highD}).%(\figref{fig:scaledSim}).
As with the 2-D simulation, the dPCA time and stimulus components do not appear well demixed (\figref[C]{fig:highD}).
In contrast, the stimulus kdPC separates the conditions and the interaction term appears qualitatively similar to the 2-D scaling case (\figref[E]{fig:highD}).
The interaction component for the non-scaled 3rd condition (red trace) is clustered around the origin, correctly indicating that this component was not scaled (i.e., the red component was the mean of the other training components).
Thus, performing demixed dimensionality reduction can extend visualization of datasets beyond PCA, and kdPCA can aid in the demixing process when components share overlapping subspaces.

 \begin{figure}[t!]
\centering
\includegraphics[width=0.75\linewidth]{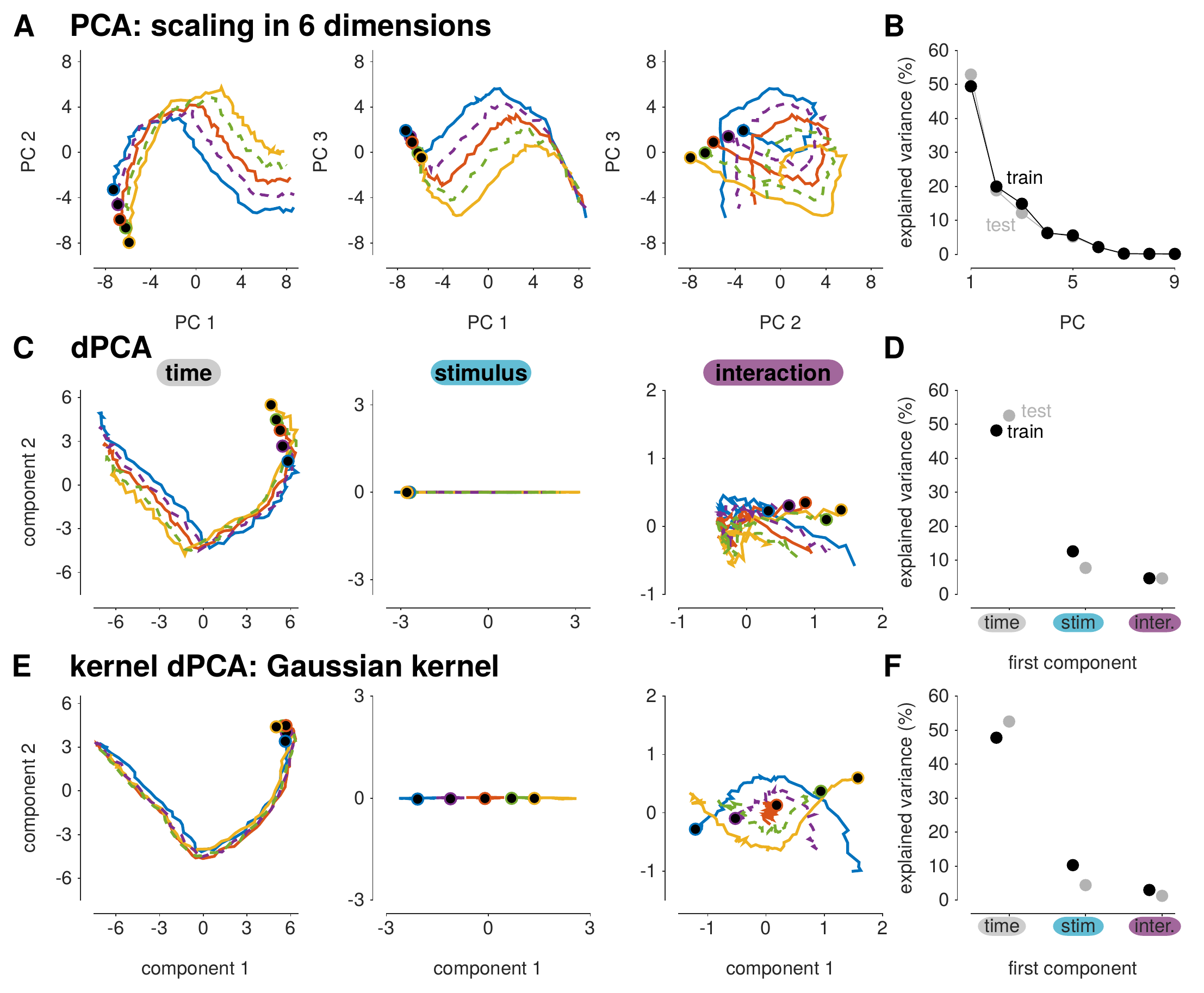}
\caption{ PCA, dPCA, and kdPCA for one random instance of the scaling simulation in six dimensions.
({\bf A}) The first three principal components. Each color indicates one condition.
PCA was fit to three training conditions (solid lines) and tested on two conditions (dashed lines).
({\bf B}) The percent variance explained by the first nine PCs for the training (black) and test (gray) components.
({\bf C}) The first two time (left), stimulus (middle), and interaction (right) demixed PCs.
({\bf D}) The variance explained in the data by the first time, stimulus, and interaction components for the training (black) and test (gray) conditions.
({\bf E-F}) Same as C-D for kdPCA. 
}
\label{fig:highD}
\end{figure}

\subsection{Data Example 1: Isolating decision-related activity in rat OFC}

I reanalyzed one of the datasets used by Kobak {\emph{et al.}} to demonstrate dPCA for extracting stimulus- and decision- related activity in a large population of neurons.
This analysis included 214 neurons from rat OFC during an odor categorization task (\figref[A]{fig:dataAll}).
In this task, the rats were presented with a mixture of two odorants.
The relative concentration of the odorants varied over trials.
The rats were trained to indicate which odorant was strongest by a nose poke into the left or right choice port.
I analyzed the trial-averaged activity under eight conditions covering four concentration levels for the two possible decisions.
The two easiest conditions (the $100\%$ concentration) in which the rats performed nearly perfectly, were used only as test conditions and were thus not used to fit dPCA or kdPCA (as was done in Kobak~\emph{et al.})
I fit dPCA and kdPCA to these data and found qualitatively similar components that capture a comparable amount of variance.
The projections onto the components are shown as a function of time in \figref[B-C]{fig:dataAll} to compare with the figures in Kobak~{\emph{et al}}.
\figref[A-B]{fig:data2D} shows the trajectories in two-dimensional component space to compare with the simulation figures.

 \begin{figure}[t!]
\centering
\includegraphics[width=0.9\linewidth]{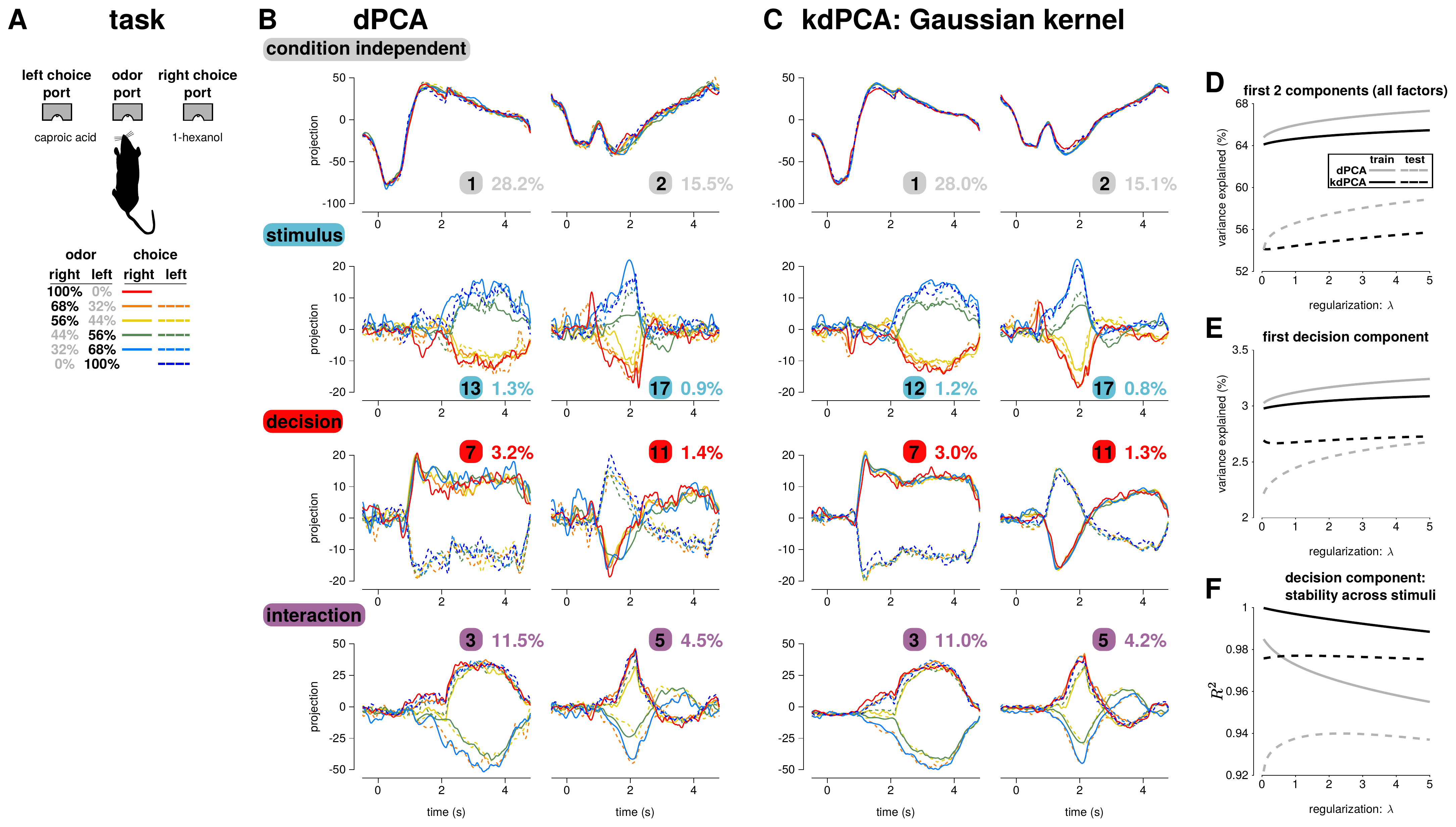}
\caption{ dPCA and kdPCA produce similar results in reducing the dimensionality of data recorded from 214 neurons in rat OFC from~\cite{KepecsEtAl2008}.
({\bf A}) The rat categorizes concentrations of two odors (caproic acid and 1-hexanol) and signals choice by selecting the left or right choice port.
({\bf B}) The average neural activity projected onto the first two condition independent, stimulus, decision, and stimulus-decision interaction dPCs.
Following Kobak~{\emph{et al.}}, the projections are shown as a function of time in the trial.
The percent variance explained and the component's rank (from most variance explained to least) is given in each plot.
({\bf C}) The same as B for the Gaussian kdPCs.
({\bf D}) The percent variance explained in the training data (solid lines) and test data (dashed lines) as a function of the regularization parameter $\lambda$ by the sum of the first two components for all factors (i.e., sum all the components shown in panels B-C) for dPCA (gray) and kdPCA (black).
({\bf E}) The percent variance explained in the training data (solid lines) and test data (dashed lines) as a function of the $\lambda$ by the first decision component (third row left of panels B-C) for dPCA (gray) and kdPCA (black).
({\bf F}) The stability factor of the first decision component for the four training stimulus conditions (solid lines) and two test conditions (dashed lines) as a function of the regularization parameter $\lambda$.
Higher values indicate less stimulus-dependence of the decision-only component.
}
\label{fig:dataAll}
\end{figure}

 \begin{figure}[t!]
\centering
\includegraphics[width=0.8\linewidth]{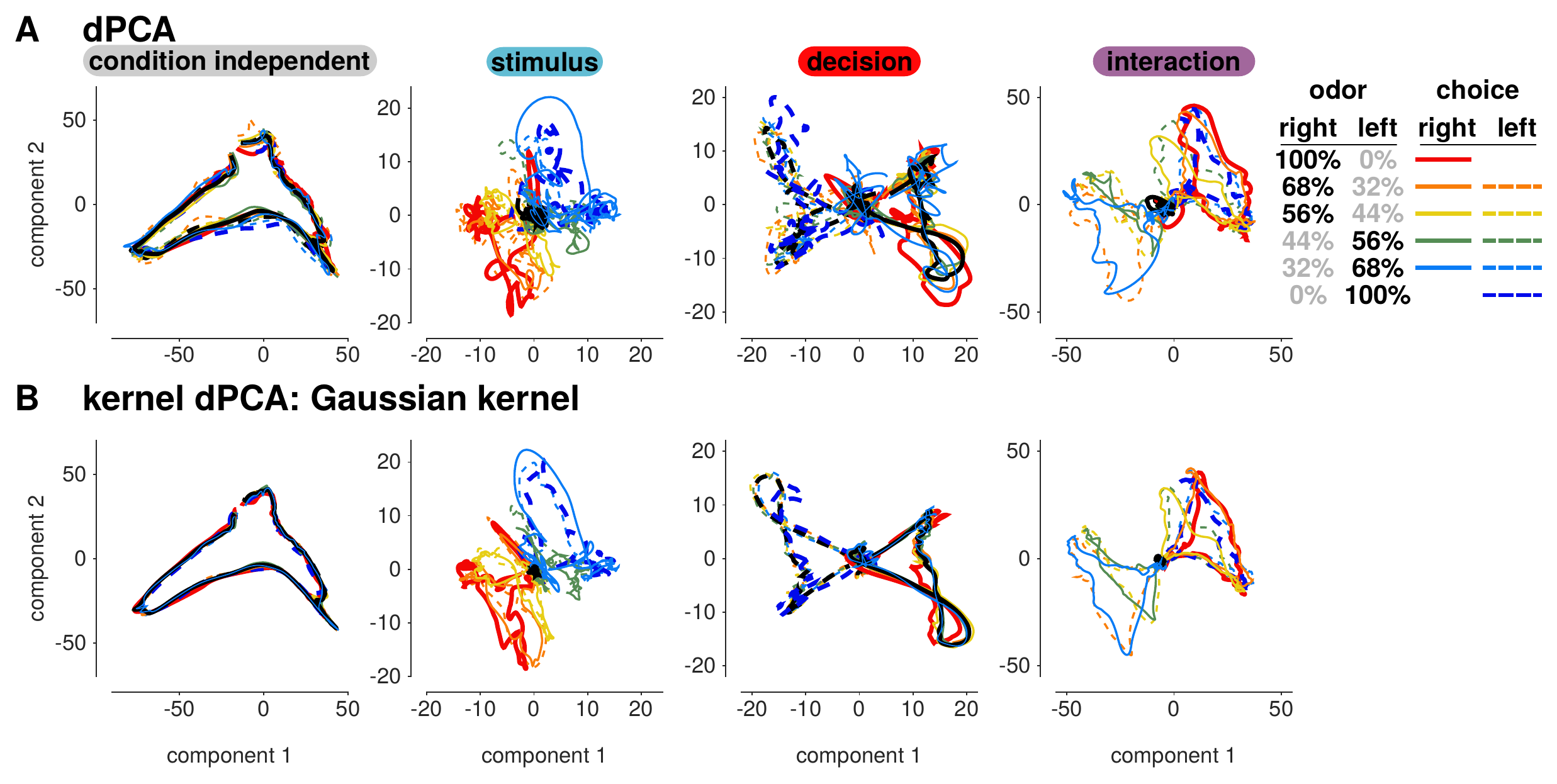}
\caption{ The dPCs ({\bf A}) and kdPCs ({\bf B}) fit to data from~\cite{KepecsEtAl2008}.
 The components are the same as in \figref[B-C]{fig:dataAll}, but the first two components are plotted together instead of as a function of time for comparison with simulation figures.
}
\label{fig:data2D}
\end{figure}

Here I focus on the ability of kdPCA to isolate (or demix) the decision-related activity from stimulus-dependent responses.
The dPCA decision components show some variability across stimuli (\figref[B 3rd row]{fig:dataAll}).
Although decisions depend on the stimulus, these dependencies should ideally be confined to the interaction term.
Thus, the stimulus and decision terms are not completely demixed by dPCA.

To quantify the stimulus-dependence in the decision components from dPCA and kdPCA, I predicted the activity projected onto the first decision component for each condition (\figref[B 3rd row]{fig:dataAll}) using the mean decision component of the other stimulus conditions from the same decision.
Stability of the decision component across stimulus conditions in the training data is quantified as the average $R^2$ between of each stimulus condition as predicted by the mean of the remaining three conditions.
Stability of the test decision component is the average $R^2$ of the two test conditions projected into the decision space as predicted by the mean projection of the four training stimulus conditions.
This metric was computed across a range of settings for the regularization parameter (\figref[F]{fig:dataAll}).
The kdPCA decision component, while similar in shape to its dPCA counterpart, shows greater stability across stimuli in both the training and test conditions over the entire range of $\lambda$.
Selecting the optimal $\lambda$ to achieve the best test set stability, dPCA's stability is $R^2_{train}=0.965$ and $R^2_{test}=0.940$ while the kdPCA training stability is $R^2_{train}=0.996$ and $R^2_{test}=0.977$.
The decision-related activity of the first dPCA explained only $0.15\%$ more variance than the kdPCA component, but kdPCA explained $1.1\%$ more of the variance in the test conditions (\figref[E]{fig:dataAll}).
Thus, kdPCA finds a more demixed representation of the neural data which shows superior generalization performance than the linear components recovered by dPCA and this improvement was not dependent on regularization.

\subsection{Data Example 2: decision making in macaque LIP}

The second example application of kdPCA to neurons recorded from macaque LIP during a visual decision-making task~\citep{MeisterHennigHuk2013}.
The stimulus consisted of a field of moving dots, and the monkey's task was to determine the direction of motion (e.g., left or right; \figref[A]{fig:lip:main}).
The difficulty of the task was controlled by adjusting the percentage of dots moving towards one of the two choice directions on each frame (coherence).
The remaining dots were randomly replotted.
On each trial, the monkey was required to fixate at a central point on the screen.
Two choice targets then appeared: one placed inside LIP cell's response field (RF) and one in the opposite hemifield.
The motion was then displayed for 500-1000~ms beginning 200~ms after the choice targets appeared.
After receiving a go signal, the monkey responded with its choice by saccading to one of the choice targets (In-RF or Out-RF choices).
This particular variant of the task included an extra condition: with a probability of $50\%$ on each trial, the choice targets were displayed for only 100~ms and the decision was reported by a memory-guided saccade (targets-FLASH trials).
On the remaining trials, the targets remained visible until after the saccade (targets-ON trials).
However, the target condition was irrelevant to the task, and the target positions remained the same throughout each session.
The animals' behavioral performance was similar across the two conditions.

Single neuron activity in LIP during the stimulus period of this task reflects motion strength and direction and animal's choice~\citep{RoitmanShadlen2002}, and decision-related activity is thought to follow low-dimensional dynamics~\citep{GanguliEtAl2008}.
The LIP responses in this dataset also depended in the target condition in a way consistent with a multiplicative gain change~\citep{ParkEtAl2014}.
That is, the firing rates in the targets-FLASH condition could be described by the rate the targets-ON condition multiplied by a scaling factor (\figref[B]{fig:lip:main}).
I therefore aimed to visualize the interaction between decision and target conditions in the population using dPCA and kdPCA.

Here, I analyzed spiking activity in a 700~ms window starting 200~ms before motion onset (choice target onset).
Motion coherence levels were grouped into high ($12.8, 25.6,$ and $51.2\%$; easy trials), low ($3.2$ and $6.4\%$; difficult trials), and 0\% coherence (ambiguous trials on which on average no motion occurred) conditions.
The resulting five stimulus levels were -high, -low, zero, +low, and +high where positive values indicates motion towards the LIP RF and negative indicates motion towards the out-RF target.
Trials were then sorted into 20 total conditions with three parameters: five motion coherence and direction, two choices (In-RF and Out-RF), and two target conditions (targets-ON or targets-FLASH).
The four 0\% coherence stimulus conditions were held as a test set.

\begin{figure}[t!]
\centering
\includegraphics[width=0.9\linewidth]{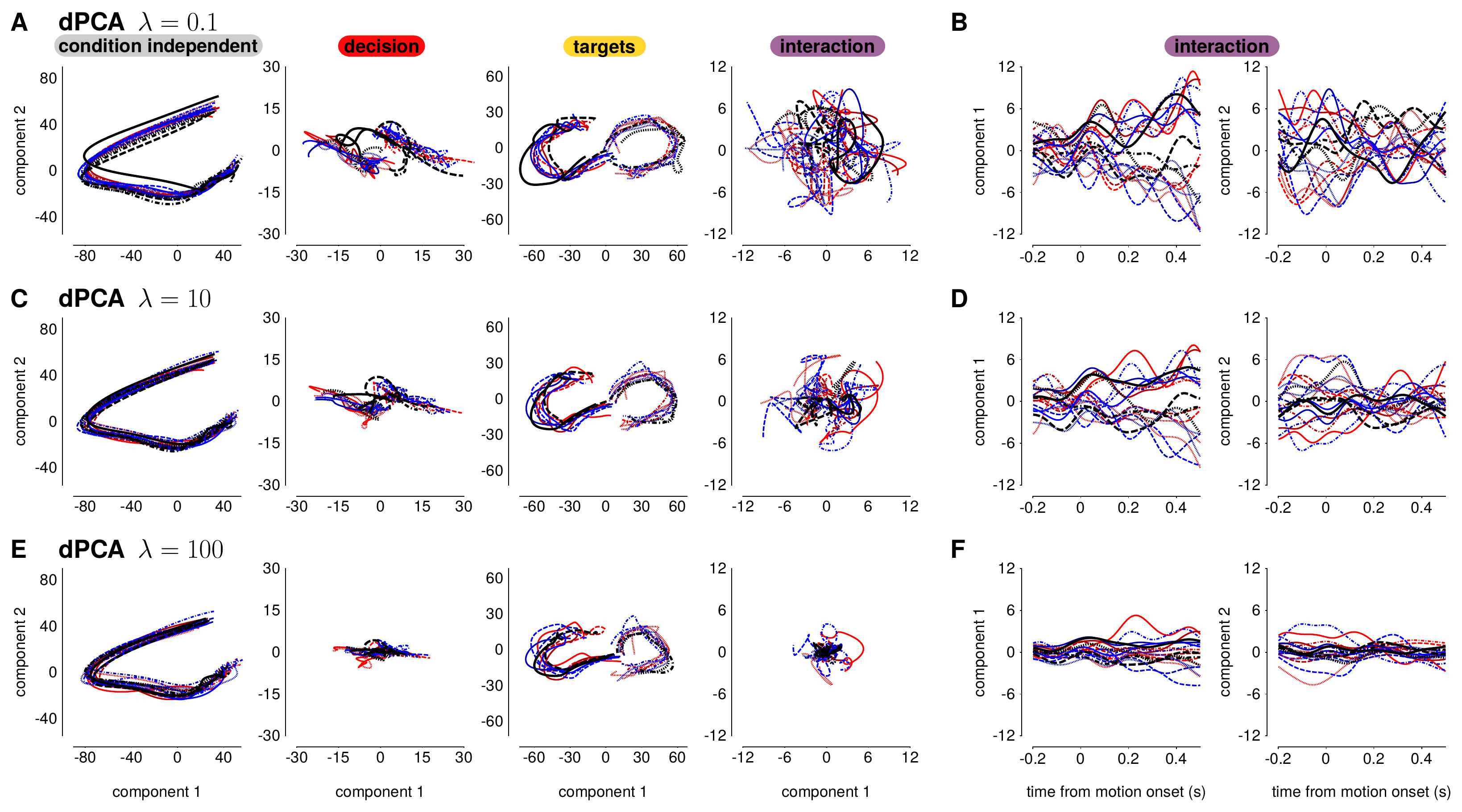}
\caption{ The dependency of the dPCs recovered for the LIP data in \figref[A-D]{fig:lip:main} on regularization.
({\bf A-B}) The dPCs fit with with regularization $\lambda=0.1$, ({\bf C-D}) $\lambda=10$, and ({\bf E-F}) $\lambda=100$.
}
\label{fig:lip:regularization}
\end{figure}

I quantified the overlap across dPCA components by computing the dot product between the first stimulus, decision, and target dPCA encoding vectors.
All three vectors were significantly non-orthogonal ($\mf_{1,stim}\cdot\mf_{1,dec} = 0.46$, $\mf_{1,stim}\cdot\mf_{1,targ} = 0.60$, $\mf_{1,dec}\cdot\mf_{1,targ} = 0.67$, $p<0.001$). %($A_{stim,dec} = 62\degree$, $A_{stim,targ} = 53\degree$, $A_{dec,targ} = 48\degree$, $p<0.001$).
As anticipated in the situation where components project into overlapping subspaces, the decision, target, and the decision-target interaction components from dPCA appear less demixed than kdPCA (\figref[C,E]{fig:lip:main}).
The average activity of LIP neurons during this task canonically ramp up with In-RF choices and down with Out-RF choices~\citep{RoitmanShadlen2002,GoldShadlen2007}.
The ramp-like structure of the first kdPCA interaction component is consistent with the difference between choice-dependent ramping at high gain (targets-ON) and low gain (targets-OFF) states (\figref[B,F left]{fig:lip:main}).
Although the interaction component in the 0\% coherence test conditions shows that the stimulus-independent components were not completely recovered by kdPCA, the multiplicative gain interaction pattern was still qualitatively consistent in the test conditions.
The second kdPCA interaction component reveals an interaction between the decision and target space that is quenched by the onset of the motion stimulus.
These structures were less visible in the dPCA components, which appear much more noisy.
Moreover, exploring a wide range of regularization values indicates that the differences between dPCA and kdPCA were not due to sensitivity in regularization for dPCA (\figref{fig:lip:regularization}).
Thus, kdPCA can aid in recovering demixed components when two experimental parameters share overlapping subspaces in the neural activity.

 \begin{figure}[t!]
\centering
\includegraphics[width=0.9\linewidth]{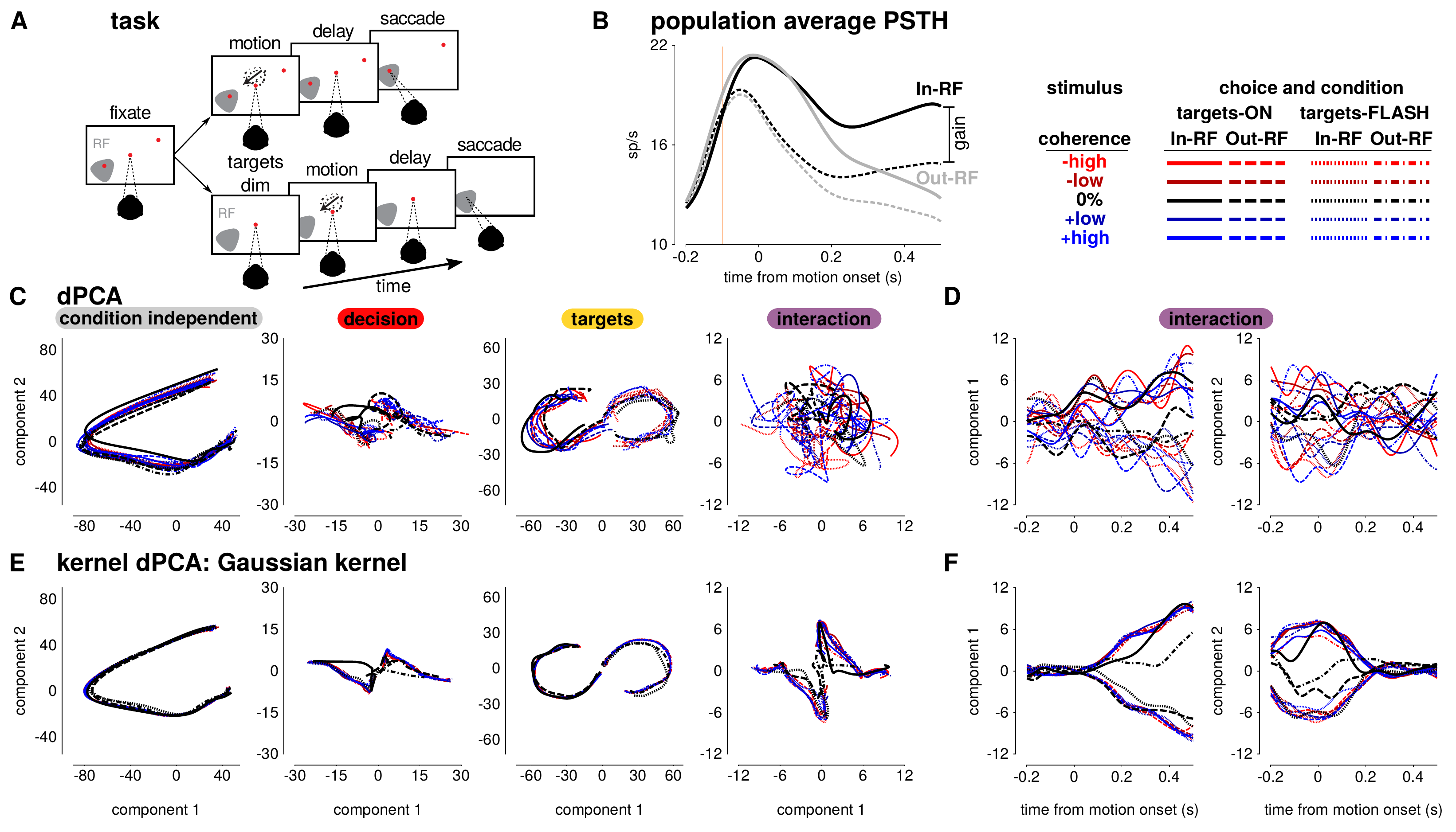}
\caption{ ({\bf A}) Diagram of the LIP motion direction discrimination task.
({\bf B}) The population average spike rate on correct trials sorted by the monkey's choice (In-RF in black and Out-RF in gray) and target condition (targets-ON solid traces, targets-FLASH dashed).
The difference between the target conditions is consistent with a multiplicative gain decrease after the targets are dimmed (denoted by line at 100~ms before motion stimulus onset).
({\bf C}) The condition independent, choice, target condition, and interaction dPCs for 69 LIP neurons.
The interaction term is between choice and target conditions only; the component is stimulus independent.
The traces show the first two components for each condition.
The stimulus strength and direction are denoted by color.
The choice and target conditions are given by the line style: solid traces for
In-RF choice and targets-ON condition, dashed traces for Out-RF choice and targets-ON condition, dotted traces for In-RF choice and targets-FLASH condition, and dot-dash traces for Out-RF choice and targets-FLASH condition.
The dPCs here were fit with regularization $\lambda = 1.0$.
The $0\%$ coherence conditions (black traces) were not used to fit the demixed components.
({\bf D}) The first two interaction components in B as a function of time.
({\bf E-F}) same as C-D for the kdPCs.
}
\label{fig:lip:main}
\end{figure}

\section{Discussion}

I have proposed an extension to dPCA that can extract nonlinear components from neural populations which are related to experimental parameters.
The resulting components provide a more demixed representation of the data when parameters and interaction effects span overlapping subspaces.

The simulations presented here explored the problem of finding relevant demixed components when the true components are scaled by a gain factor or rotated under different conditions.
The dPCA components recovered from the simulated activity were less associated with the experimental parameters of interest than the kdPCA components.
In the scaling and rotations simulations, dPCA recovered the original mixed dimensions, and thus was unable to go significantly beyond standard PCA.
Incomplete demixing could potentially bias estimates of the variance explained in the data by an individual parameter, because the variance explained by a poorly demixing component does not solely reflect the parameter of interest.
This may account for the decrease in variance explained by the kdPCA components compared to the dPCA components (\tableref{tab:perVarExplained}).
In contrast, kdPCA successfully recovered demixed stimulus and time components.

In data recorded from rat OFC during a decision-making task, I found that kdPCA could extract decision-dependent components that were more independent across stimulus conditions than dPCA.
Additionally, the nonlinear demixed components generalized more successfully to test conditioned than the linear components.
For neurons recorded in macaque LIP during a visual motion discrimination task, kdPCA recovered interaction terms that were consistent with gain changes in the population activity caused by a nuisance variable (targets that were either held on or flashed).
These results suggest that nonlinear mixing was present between parameters in these data sets. 

The form of kdPCA presented here considered only nonlinear dimensionality reduction in the decoding stage of dPCA.
Although nonlinear components can be more difficult to interpret than linear components, the variable-dependent reconstructions in this formulation must sum together to approximate the complete neural response.
As a result, stimulus-dependent scaling and rotation could still be visualized in the interaction terms recovered by Gaussian kdPCA in the simulations explored here.
However, future work on kernelized extensions to dPCA could consider either the encoding or marginalization steps.
Nonlinear methods that capture both the encoding and decoding stages could aid in discovering activity that lies along low-dimensional manifolds embedded in higher dimensional spaces~\citep{MishneEtAl2017}.
Kernelized marginalization could potentially be accomplished by performing kernel-based dimensionality reduction, such as kPCA, on the original data and then apply dPCA or kdPCA on the data projected onto the first several components. 
However, including a nonlinear projection to either the encoder or marginalization steps risks lowering interpretability of the demixed components across experimental parameters because the components would no longer sum together in neural firing rate space.

The flexibility of kdPCA requires tuning more free parameters than dPCA. 
In addition to selecting regularization terms, the user must also chose an appropriate kernel, and select any kernel parameters (e.g., bandwidth).
The same cross-validation procedure recommended by Kobak {\emph{et al.}} could be applied to kdPCA when considering trial-averaged responses.
However, cross-validation techniques for demixed dimensionality reduction remain an open direction for future research.

Visualizing the complex responses of large neural populations to yield insights about neural dynamics and processing will require nonlinear dimensionality reduction techniques~\citep{StopferEtAl2003,CunninghamYu2014, GallegoEtAl2017}.
kdPCA is one such tool that may aid investigators seeking to unravel neural manifolds.

\section*{Acknowledgments}
I thank David Freedman, Jonathan Pillow, Leor Katz, Benjamin Scholl, and Jacob Yates for their helpful comments and discussion, and Alex Huk and Miriam Meister for sharing LIP data.
I am grateful for the encouraging and highly constructive comments from the reviewers at NBDT.

% Submissions are not required to reflect the precise reference formatting of the journal (use of italics, bold etc.), however it is important that all key elements of each reference are included.
\bibliography{kdPCAbib}

\end{document}